\documentclass[pdflatex,sn-mathphys-num]{sn-jnl}

\usepackage{graphicx}%
\usepackage{multirow}%
\usepackage{amsmath,amssymb,amsfonts}%
\usepackage{amsthm}%
\usepackage{mathrsfs}%
\usepackage[title]{appendix}%
\usepackage{xcolor}%
\usepackage{textcomp}%
\usepackage{manyfoot}%
\usepackage{booktabs}%
\usepackage{algorithm}%
\usepackage{algorithmicx}%
\usepackage{algpseudocode}%
\usepackage{listings}%

\newcommand{\Fl}{\mathcal{F}}
\newcommand{\V}{\mathcal{V}}

\theoremstyle{thmstyleone}%
\theoremstyle{thmstyletwo}%

\theoremstyle{thmstylethree}%

\begin{document}
\title[Universal scalings and switching entropy in yield-stress fluids]{Universal scalings and switching entropy in yield-stress fluids}

\author[1]{\fnm{Rajam} \sur{Elancheliyan}}\email{rajam.elancheliyan@gmail.com}

\author[1]{\fnm{Jean Marc} \sur{Fromental}}\email{jean-marc.fromental@umontpellier.fr}

\author[1]{\fnm{Edouard} \sur{Chauveau}}\email{edouard.chauveau@umontpellier.fr}

\author*[1]{\fnm{Domenico} \sur{Truzzolillo}}\email{domenico.truzzolillo@umontpellier.fr}

\affil*[1]{\orgdiv{L2C}, \orgname{Univ Montpellier, CNRS, F-34095 Montpellier, France}, \orgaddress{\city{Montpellier}, \postcode{F-34095}, \country{France}}}




\abstract{

Yield-stress fluids transition from solid-like to liquid-like behavior at a critical stress threshold, governing phenomena from industrial processing to geological flows. While predominantly investigated under steady shear, large-amplitude oscillatory tests force these materials to cyclically navigate between arrested and fluidized states. Here, we discover a hidden universal behavior where the dynamic viscoelastic moduli of yield-stress fluids collapse onto master curves, revealing that these materials rearrange almost instantaneously to maintain a constant stress state. We fully capture this behavior using a novel theoretical framework based on the minimization of a governing function that exhibits symmetry breaking. Our findings reveal that recoverable elastic energy, yielding abruptness, and entropy production during stress inversion are fundamentally intertwined. This connection provides a unified physical picture for the dynamic yield stress, offering a novel thermomechanical foundation to define and predict this threshold across soft matter physics and materials science.

}




\maketitle
\section*{Introduction}
Yield-stress fluids, ranging from colloidal suspensions to cooled lava, emulsions and foams, are defined by their dual nature: they behave as elastic solids at rest but flow as liquids when subjected to stresses exceeding a characteristic threshold \cite{bonnYieldStressMaterials2017, mewisColloidalSuspensionRheology2013, coussotRheometryPastesSuspensions2010,griffithsDynamicsLavaFlows2000}. Specifically, this transition is governed by the dynamic yield stress, below which the material inevitably undergoes structural arrest \cite{bonnYieldStressMaterials2017}. Numerous efforts \cite{bocquetKineticTheoryPlastic2009, kamaniUnificationRheologicalPhysics2021, sethMicromechanicalModelPredict2011, voigtmannYieldStressesFlow2011, sollichRheologySoftGlassy1997, ikedaUnifiedStudyGlass2012, fuchsTheoryNonlinearRheology2002} have attempted to couple elastic and dissipative phenomena to describe the yielding transition. Despite such work, the interplay governing both the magnitude of this threshold and how abruptly an arrested state transitions to full strain-induced fluidization has yet to be established, limiting our ability to treat the yield stress as a true thermomechanical property rather than a mere fit parameter \cite{nicolasDeformationFlowAmorphous2018}.

Crucially, a predictive bridge between linear response properties—such as the shear modulus—and the stress required to fluidize the system remains elusive. Likewise, it is still exceedingly difficult to map the material's ductility or brittleness onto the precise yielding threshold, preventing any straightforward inference of large-deformation behavior from deformation-independent quantities.

In this context, Large-Amplitude Oscillatory Shear (LAOS) effectively probes this transition, forcing the material to cyclically navigate yielding by probing both solid-like recovery and flow-induced rejuvenation within a single period \cite{donleyElucidatingOvershootSoft2020, aimeUnifiedStateDiagram2023a}. Crucially, in the low-frequency, large-amplitude limit, the response becomes asymptotically equivalent to steady shear, mapping the distinct steady states characteristic of the material's flow curve.

Here, we show that the yielding of soft glassy materials is governed by a universal scaling of the first-harmonic viscoelastic moduli, unveiling that yield-stress fluids act as bistable oscillators transitioning between non-equilibrium steady states and controlled by a time-dependent relaxation rate. By introducing a fluidity model based on a frozen-time Lyapunov function \cite{khalilNonlinearSystemsHauptbd2002, naserStabilityClassSlowly2018} — a mathematical function evaluating instantaneous stability under slow driving and acting as a free energy to be minimized - we link yielding to three core parameters: the shear modulus, the viscoplastic fragility defining yielding abruptness, and a novel ``switching entropy'' produced during stress inversion. 
This framework categorizes diverse complex fluids through a single scalar---the plasticity production exponent, $\lambda$---redefining the dynamic yield stress through the ratio between recoverable elastic energy and the dissipative cost of structural reconfiguration.
\section*{Results}
\subsection*{Universal scalings}
\begin{figure}
    \centering
    \makebox[\textwidth][c]{
    \begin{minipage}{1.2\textwidth}
    \centering
    \includegraphics[width=\linewidth]{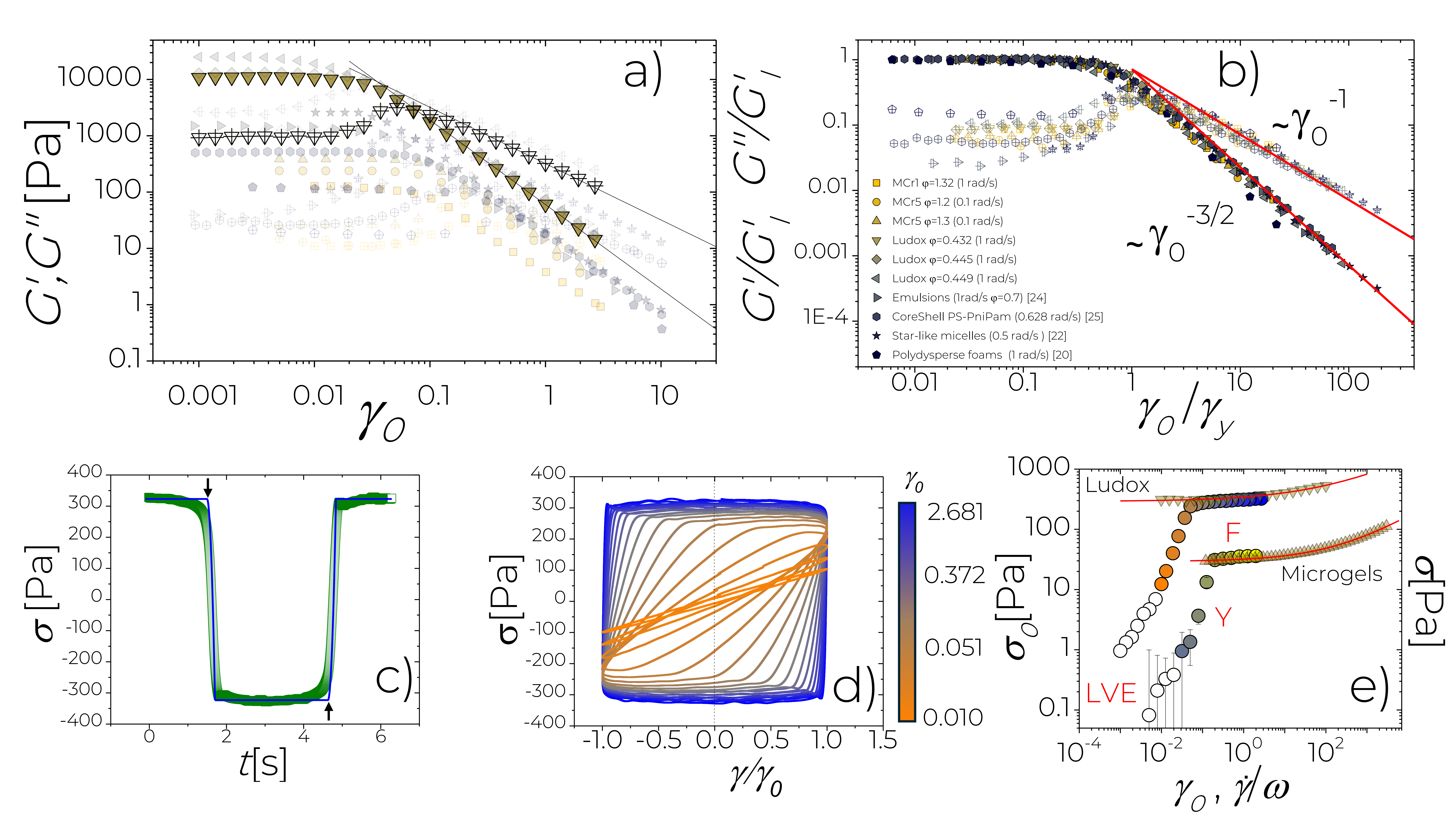}
    \caption{\textbf{Universal scalings in yield stress fluids (Experiments)}. 
    \textbf{a)} Oscillatory dynamic strain sweep experiments. First-harmonic moduli $G'(\gamma_0)$ and $G''(\gamma_0)$ at fixed frequencies for all the systems listed in the legend of panel b). The highlighted data points refer to the glassy suspension of Ludox TM50 whose intra-cycle stress response is shown in panel d). The solid lines are the best fits of the power laws $G'=K_1\gamma_0^{-\frac{3}{2}}$, $G''=K_2\gamma_0^{-1}$ to the data in the fluidized regime ($0.13\leq\gamma_0\leq 2.68$). Here $K_1$ and $K_2$ are empirical constants. The remaining data, are shown with partial transparency for clarity. \textbf{b)} Scaled first-harmonic moduli $\left(G'/G'_l,\,G''/G'_l\right)$ \textit{vs} $\gamma_0/\gamma_y$ for all the data shown in panel a). The red solid lines are the universal decays predicted by the elastic-recoil/steady stress (ERSS) approximation discussed in the main text and SI. Data for emulsions, core-shell colloids, star-like micelles and polydisperse foams were digitized, re-analyzed and adapted from Refs. \cite{masonYieldingFlowMonodisperse1996}, \cite{carrierNonlinearRheologyColloidal2009},\cite{poulosLargeAmplitudeOscillatory2015}, \cite{marmottantElasticPlasticViscous2007}. \textbf{c)} Single stress cycle $\sigma(t)$ for Ludox TM50 ($\varphi=0.432$) at $\omega=1$ rad/s and $\gamma_0=$2.68 (green points). Solid blue lines show the ERSS approximation: the stress over a cycle is described as a sequence of elastic recoils $\sigma(\tau)\simeq\sigma_r(\tau)=\frac{1}{2}\omega^2G_c\gamma_0\tau^2-\sigma_p$ following strain rate reversal and intervals of constant stress $\sigma(\tau)=\sigma_p$, where $\tau=t-t_0$ is the time elapsed since the shear rate changed sign, and $G_c$ is the low-frequency (cage) modulus of the specific system. Vertical arrows point to the $\dot{\gamma}$ reversal.
    \textbf{d)} Intra-cycle stress response $\sigma(\gamma/\gamma_0)$ for the Ludox TM50 suspension ($\varphi=0.432$) at $\omega=1$ rad/s. \textbf{e)} Stress at zero strain (circles) $\sigma_0(\gamma_0)$ for jammed microgels MCr5 ($\varphi=1.3$) and Ludox TM50 ($\varphi=0.432$). The flow curves (triangles) under steady rate as a function of the scaled rate $\dot{\gamma}/\omega$ are also shown for the same systems for comparison, where $\omega$ represents the oscillation frequency of the corresponding oscillatory tests. The red lines are the best Herschel-Bulkley fits to the steady rate data.
    }
    \end{minipage}
    }
    \label{fig1}
\end{figure}

We performed dynamic strain sweep experiments under oscillatory strain $\gamma(t)=\gamma_0\sin(\omega t)$, alongside steady-state flow curve measurements, across a diverse set of glassy Ludox TM50 silica suspensions and jammed PNIPAM microgels (coded as MCr1 and MCr5) showing yield stress behavior (extended data in SI). The stress waveforms were extracted and analysed in detail for two of these systems: jammed microgels (MCr5) at an effective volume fraction \cite{philippeGlassTransitionSoft2018c} $\varphi=1.3$ and a glassy Ludox TM50 suspension at $\varphi=0.432$. The first-harmonic moduli, $G'(\gamma_0)$ and $G''(\gamma_0)$, for the latter are highlighted in figure~\ref{fig1}-a, alongside a comprehensive collection of glassy and jammed systems—including emulsions, star-like micelles, polydisperse foams and core-shell PS-PNIPAM microgels — all probed at frequencies $\omega \leq 1$~rad/s. 
Remarkably, when rescaled by the linear storage modulus $G'_l$, the moduli of all these systems---both investigated here and compiled from existing literature---collapse onto two distinct master curves above yielding (Fig.~\ref{fig1}-b). In this regime, they exhibit power-law decays with exponents closely following $G' \sim \gamma_0^{-3/2}$ and $G'' \sim \gamma_0^{-1}$ (SI).

The exponents characterizing the moduli decay have been debated within the framework of mode-coupling theory \cite{miyazakiNonlinearViscoelasticityMetastable2006}, soft glassy rheology \cite{sollichRheologicalConstitutiveEquation1998} and elastoviscoplastic models \cite{marmottantElasticPlasticViscous2007}, while experimental studies have reported a wide dispersion of values \cite{koumakisDirectComparisonRheology2012, marmottantElasticPlasticViscous2007,poulosLargeAmplitudeOscillatory2015}, leaving a universal consensus elusive. Nonetheless, experiments \cite{koumakisDirectComparisonRheology2012} suggest that universal exponents might emerge for any soft system, provided the colloidal volume fraction is sufficiently high—as also documented for concentrated star-like micelles \cite{poulosLargeAmplitudeOscillatory2015} and polydisperse foams \cite{marmottantElasticPlasticViscous2007,saint-jalmesVanishingElasticityWet1999}— together with a progressive weakening of the strain-amplitude dependence of the first-harmonic shear stress, as shown for emulsions \cite{masonYieldingFlowMonodisperse1996}.

Here we first establish that the exponent pair $(-3/2,\,-1)$ is a robust signature of yield stress fluids at low frequencies, where the stress plateau observed under oscillatory strain is invariant with respect to strain amplitude. An elastic-recoil plus steady-stress (ERSS) approximation predicts these exponents exactly (details in Fig. \ref{fig1} and SI):


\begin{equation}\label{scaledG'}
\frac{G'}{G_c}=\frac{9}{4\pi}\left(\frac{\gamma_y}{\gamma_0}\right)^{3/2} + O\left(\frac{1}{\gamma_0^{5/2}}\right)
\end{equation}
\begin{equation}\label{scaledG''}
\frac{G''}{G_c}=\frac{9}{4\pi}\frac{\gamma_y}{\gamma_0} + O\left(\frac{1}{\gamma_0^2}\right),
\end{equation}
where $G_c\approx G'_l$ is the low-frequency modulus (SI) of the specific system, $\gamma_y=\frac{16}{9}\frac{\sigma_p}{G_c}$ is the strain at which the two power laws intersect, and $\sigma_p$ is the constant stress attained during a strain cycle (see Fig. \ref{fig1}-c,d).
These scaled moduli decays are overlaid to the experimental ones in figure \ref{fig1}-b. The same scaling behavior, albeit characterized by different prefactors, was previously proposed by Marmottant et al. \cite{marmottantElasticPlasticViscous2007} using a Kelvin-Voigt model in the more specific context of the non-linear rheology of liquid foams.\\ 
We are thus presented with a significant result: for this scaling to hold, the plateau stress $\sigma_p$ must be independent of the amplitude.\\ 
The evolution of the stress $\sigma(\gamma)$ for the glassy Ludox at $\varphi=0.432$ at increasing amplitudes is shown in figure~\ref{fig1}-d. 
The stress at zero strain, $\sigma_0$, a proxy for the steady-state stress, is reported in figure~\ref{fig1}-e for both the Ludox and the MCr5 microgel suspensions. As a function of $\gamma_0$, $\sigma_0$ exhibits three regimes: a linear response regime (LVE) with $\sigma_0=G''_{l}\gamma_0$, where $G''_{l}$ is the loss modulus in the linear regime; a superlinear yielding (Y) regime; a high-amplitude fluidized (F) regime in which $\sigma_0$ becomes nearly independent of strain amplitude. Notably, this constant stress matches well the steady stress measured under continuous shear (triangles in Fig.~\ref{fig1}-e) within experimental uncertainties, pointing to a very rapid re-adaptation of the characteristic relaxation rate of the systems to the imposed oscillatory deformation.\\
\subsection*{The bistable fluidity framework (BFF)}
Inspired by the observed universal behavior of the moduli and the existence of a steady stress under oscillatory shear, we develop a model based on the minimization of a Lyapunov function — treated in a frozen-time formalism \cite{khalilNonlinearSystemsHauptbd2002,naserStabilityClassSlowly2018} — which generates the equation of motion for the material's relaxation rate, fits the full intracycle stress response of yield stress fluids, recovers the observed experimental scaling and unveil a novel relationship between yield stress, entropy production and abruptness of yielding transition.\\ 
In the theory of dynamical systems, a Lyapunov function serves as a generalized scalar potential that strictly decreases along the system's trajectories, akin to energy dissipating toward a stable minimum. By identifying such a governing function for a complex fluid, one can guarantee that the material’s internal state relaxes toward a well-defined "equilibrium". Here, we exploit this framework to drive the time evolution of the material's relaxation rate, integrating it naturally into a Maxwell-like constitutive picture. We thus introduce a dynamical framework based on an energy-like minimization. By employing a frozen-time formalism — which treats the rheological state of the material defined by strain, strain rate and stress as temporarily fixed during the oscillations — our model naturally generates the equation of motion for the material's relaxation rate. 

We thus adopt a generalized Maxwellian framework to capture the universal mechanical response of yield stress fluids. This constitutive backbone allows us to isolate the fundamental competition between elastic stress accumulation and complex relaxation dynamics. Under oscillatory shear, this competition arises from the balance between the elastic recoil associated with structural recovery and plastic fluidization, effectively mediated by a time-dependent relaxation rate—historically termed ``fluidity'', $\Fl(t)$ \cite{bocquetKineticTheoryPlastic2009, derecRheologyAgingSimple2001,derecAgingNonlinearRheology2003}.\\
Our central claim is therefore that the low-frequency dynamics are governed by the evolution of $\Fl(t)$, which is coupled to the stress, strain, and their respective time derivatives via a generic functional $\Psi$:
\begin{flalign}
& \dot{\sigma}(t)=G_c\dot{\gamma}(t)-\sigma(t)\left[\Gamma_0(\omega)+\Fl(t)\right] && \label{fluidity1}\\
& \dot{\Fl}(t)=\Psi(\sigma(t),\gamma(t),\dot{\sigma}(t),\dot{\gamma}(t)), \label{fluidity2}&&
\end{flalign}
where $\Gamma_0(\omega)$ is a frequency dependent mechanical relaxation rate that has to be introduced to set the linear viscoelastic (LVE) response of the system.
In this framework, the inclusion of the fluidity $\Fl(t)$, due to plastic events, adds a purely dissipative channel into the Maxwell model (SI). 
The ratio $\sigma \Fl / G_c$ governs the rate of plastic strain ($\gamma_p$), since $\sigma \Fl / G_c=d\gamma_p/dt$ (SI). 
The experimental observation of a constant stress plateau during the imposed strain cycles suggests that the dynamical system defined by \eqref{fluidity1} and \eqref{fluidity2} must possess fixed (stationary) points characterized by a stationary stress. As evidenced by Eq. \eqref{fluidity1}, under the assumption of a fully fluidized system ($\Fl \gg \Gamma_0$), this stationarity extends to the ratio between the fluidity and the applied strain rate: the steady stress condition is reached when $\frac{\Fl(t)}{\dot{\gamma}}=\frac{G_c}{\sigma_p}$. 
Hence, the existence of a steady stress under oscillatory forcing imposes that fluidity acts as a fast variable — readjusting almost instantaneously to changes in $\gamma$ and $\dot{\gamma}$ — allowing both strain rate and stress to be treated as quasi-static, or "frozen", within the equation of motion for $\Fl$. This separation of timescales reduces the problem to a one-dimensional manifold: the evolution of $\Fl$ occurs in a regime where all viscoelastic-plastic parameters are effectively constant, while $\Fl$ adapts instantaneously to the local state. 
We thus adopt a frozen-time formalism \cite{khalilNonlinearSystemsHauptbd2002,naserStabilityClassSlowly2018}. Given the one-dimensional nature of the problem, the dynamics can be expressed as a gradient flow — a form consistent with previous descriptions of fluidity dynamics \cite{benziUnifiedTheoreticalExperimental2019,benziContinuumModelingSoft2023}. Requiring the existence of fixed points in the space defined by the variable $\theta = \frac{\Fl}{\dot{\gamma}}$, we write the equation of motion for $\Fl$ as
\begin{flalign}
& \dot{\Fl}=-\mu(\dot{\gamma})\frac{\delta\V[\theta(\Fl)]}{\delta\Fl}\label{fluidity3},&&
\end{flalign}
where $\mu(\dot{\gamma})$ is a mobility coefficient associated to the gradient flow for $\Fl(t)$ and $\mathcal{V}(\theta)$ a suitable Lyapunov functional. In the following, we consider a spatially uniform mean field scenario; consequently, the functional $\mathcal{V}$ reduces to a standard scalar function of $\theta$, independent of spatial coordinates. 
Following Eq. \ref{fluidity3}, we adopt a Landau-like expansion for $\mathcal{V}[\theta(\Fl)]$. Since it is a function describing non-equilibrium steady states and departure from equilibria at rest, it may incorporate non-analytical terms \cite{aronLandauTheoryNonequilibrium2020,aronNonanalyticNonequilibriumField2020,ptaszynskiNonanalyticLandauFunctionals2025}, particularly in the present context of modeling the non-linear response of elastoviscoplastic systems \cite{bocquetKineticTheoryPlastic2009,benziContinuumModelingSoft2023}.

$\mathcal{V}[\theta(\Fl)]$ must be even under strain rate reversal, i.e., invariant with respect to the sign of $\dot{\gamma}$ ($\mathcal{Z}_2$ parity) and include a stabilizing term that drives the fluidity back to zero as the forcing becomes vanishingly small. Conversely, a destabilizing term---driven by the plastic strain production---is required to model the cascade of localized fluidization events \cite{mayrActivationEnergyShear2006, bouchbinderSheartransformationzoneTheoryLinear2011} within the material. We write therefore $\mathcal{V}[\theta(\Fl)]$ as 

\begin{flalign}
& \V[\theta(\Fl)]=\frac{1}{2}\alpha\theta^2-\beta|\sigma\theta|^{\lambda}\label{potential}.&&
\end{flalign}

In Eq. \eqref{potential} $\alpha$ represents the restoring constant of the potential, defining its harmonic curvature. It defines the system's resistance to the accumulation of plastic strain and it can be explicitly mapped onto the elastoviscoplastic constants of the material (SI).
In the following, the coefficient $\beta$ is set to unity, as it can be reabsorbed into the magnitude of $\alpha$ or $\mu(\dot{\gamma})$ and does not affect the temporal evolution of $\Fl$ set by eq. \ref{fluidity3}.

Regarding the exponent $\lambda$ in Eq. \eqref{potential}, stability requirements constrain its values to the range $0 < \lambda \le 2$.

\begin{figure}
    \centering
    \makebox[\textwidth][c]{
    \begin{minipage}{1.4\textwidth}
    \centering
    \includegraphics[width=1.0\linewidth]{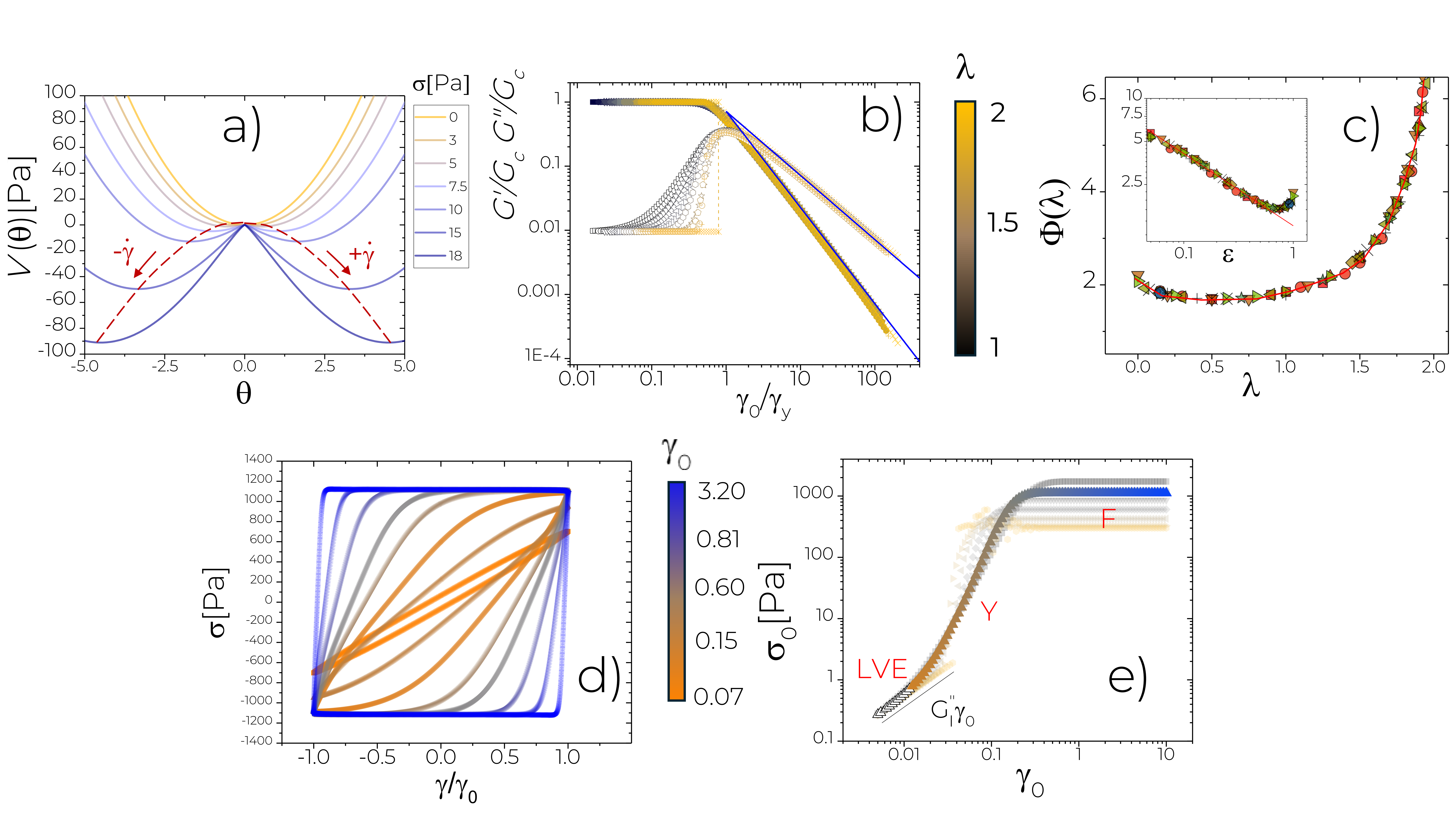}
    \caption{\textbf{Universal scalings in yield stress fluids (Theory)}. \textbf{a)} 
    Evolution of the frozen-time Lyapunov function $\mathcal{V}(\theta)$ for varying stress values as indicated in the panel (with $\alpha = 15$, $\beta = 1$, and $\lambda = 1.25$). For non-zero stress, $\mathcal{V}(\theta)$ develops a double-well profile; this symmetry-breaking mechanism drives the system into one of the two local minima depending on the sign of the shear rate $\dot{\gamma}$ (red dashed lines and arrows).\textbf{b)} Scaled first-harmonic moduli simulated according to the coupled equations \eqref{fluidity1}-\eqref{fluidity5}, for $\lambda$ varying between 1 (black) and 2 (dark yellow) and $G_c=10000$ Pa, $\sigma_c=300$ Pa, $r=150$ s$^{-1}$, $\Gamma_0=$0.0067 s$^{-1}$, $\omega=0.7$ s$^{-1}$. Solid blue lines are the moduli predicted by the ERSS equations \eqref{scaledG'} and \eqref{scaledG''}. 
    \textbf{c} Viscoplastic fragility $\Phi(\lambda)$ for different viscoelastoplastic parameters as reported in Figure S6 (SI). The red line is a guide for the eye. The inset shows the same data as a function of the reduced plastic activity exponent $\varepsilon=\frac{\left|2-\lambda\right|}{2}$. The red line is the best power-law fit to the data (see text) in the critical region ($\lambda>$1). \textbf{d)} Full stress intracycle response $\sigma(\gamma/\gamma_0)$ for the same parameters employed to compute the DSS curves shown in panel b) and $\lambda=$1.25. \textbf{e)} Stress at zero strain $\sigma_0(\gamma_0)$ extracted from the full simulated intra-cycle response for the same parameters and $\lambda$-values used to compute the first-harmonic moduli shown in panel b). The highlighted curve (orange/blue up-triangles) shows $\sigma_0(\gamma_0)$ obtained for the same parameters used in panel d) (same color code). The remaining curves, are shown with partial transparencyfor clarity.}\label{fig2}
    \end{minipage}
    }
\end{figure}

The introduction of the non-analytic term $-|\sigma\theta|^\lambda$ in Eq. \eqref{potential} effectively transforms the harmonic equilibrium at $\theta=0$ into a singular repellor for $|\sigma|>0$ (Fig. \ref{fig2}-a). This construction provides a framework to model the marginal stability of the system, capturing the instantaneous loss of stability and the subsequent structural reorganization characteristic of jammed or glassy states under infinitesimal external stress \cite{mullerMarginalStabilityStructural2015a,kamaniUnificationRheologicalPhysics2021}.

Since $\sigma\theta=G_c\frac{d\gamma_p}{d\gamma}$ (SI), $\lambda$ characterizes the rate at which plasticity is generated under given $\dot{\gamma}$ and $\sigma$ and quantifies the "abruptness" of the plastic onset; for this reason, we refer to $\lambda$ as the \textit{plasticity production exponent}.

$\mathcal{V}[\theta(\Fl)]$ develops a double-well shape as soon as a non-zero stress is applied; the corresponding minima progressively deepen until they reach a steady-state value, dictated by the coupling with Eq. \eqref{fluidity1}. 

We thus denote this approach as the \textit{Bistable Fluidity Framework} (BFF). At its core, the BFF treats the oscillatory yielding transition as a dynamic switch between two stable fixed points of $\mathcal{V}[\theta(\Fl)]$. The corresponding bifurcation diagrams are reported in SI. 

To impose that no fluidity is generated at rest and to regularize Equation~\ref{potential} for $\dot{\gamma} \to 0$, we require that the mobility coefficient $\mu$ associated with the gradient flow vanishes in the absence of external forcing. To maintain the required symmetry, $\mu(\dot{\gamma})$ must be an even function of the strain rate, leading to the expansion $\mu = \mu_2 \dot{\gamma}^2 + \mathcal{O}(\dot{\gamma}^4)$. From Eqs. \ref{fluidity3} and \ref{potential} we thus obtain the equation of motion for the fluidity to the leading order,
\begin{flalign}
& \dot{\Fl}=-r\Fl\left[1-\psi(\sigma,\sigma_c,\dot{\gamma},\Fl,\lambda)\right]\label{fluidity5}, &&
\end{flalign}
where $\psi(\sigma,\sigma_c,\dot{\gamma},\Fl,\lambda)=\left|\frac{\sigma}{\sigma_c}\right|^{\lambda}\left|\frac{\dot{\gamma}}{\Fl}\right|^{2-\lambda}$, $\sigma_c=\left(\alpha/\lambda\right)^{1/\lambda}$ is a characteristic stress arising from the minimization of $\V[\theta(\Fl)]$ with respect to $\Fl$ and $r=\mu_2\alpha$ is a dumping factor. It is worth noting that in the limit of large $r$ ($r \to \infty$), the solution of the differential equation \eqref{fluidity5} coincides—according to Tikhonov's theorem \cite{tichonovDifferentialEquations1985}—with the solution of the algebraic equation $\psi(\sigma,\sigma_c,\dot{\gamma},\Fl,\lambda) = 1$. This represents a tracking solution for the problem—that is, a solution where all derivatives vanish, yet only instantaneously as the system adiabatically follows the evolving constraints.
Crucially, this tracking solution identifies the non-equilibrium steady-state (NESS) trajectory of the material's relaxation time, corresponding to the instantaneous stress plateau. Together, the constant-stress condition and the tracking solution define a low-dimensional attractor in the $(\sigma, \Fl)$ phase space, toward which the material's dynamic state rapidly converges.\\\\
Driven by experimental evidence that a steady-state stress $\sigma_p$ is sustained under oscillatory strain, we deduce that jammed or glassy yield-stress fluids operate in the overdamped limit. 
This leads us to conclude that the dumping rate $r$ is much larger than the driving frequency $\omega$.
In this limit the solution of equation \eqref{fluidity5} under steady stress is (details in SI): 
\begin{flalign}\label{fl_alpha2red_b}
&\Fl(t)=\sqrt{\frac{1}{1+\frac{\omega^2}{r^2}}}\left(\frac{G_c}{\sigma_c}\right)^{\frac{\lambda}{2}}|\dot{\gamma}|.&&
\end{flalign}
This shows that, under steady-state stress conditions, the fluidity converges to a trajectory proportional to the strain rate. By substituting this expression into Eq. \eqref{fluidity1} and neglecting the correction due to the finite $\omega/r$ term, we find that the stationary, $\gamma_0$-independent stress is given by
\begin{flalign}
&\sigma_p = \sigma_c^{\lambda/2} G_c^{1-\lambda/2}.&&
\end{flalign}
More generally, the coupled equations \eqref{fluidity1}-\eqref{fluidity5} can be solved numerically (Methods) and both fluidity and stress computed. The resulting scaled moduli $G'(\gamma_0)/G_c$ and $G''(\gamma_0)/G_c$ are shown in figure \ref{fig2}-b for varying $\lambda$ in the range $1 \leq\lambda\leq 2$, i.e. where $\V$ is derivable at $\theta=0$. A broader set of curves, including those for $\lambda < 1$, and a short discussion of the physical relevance of $\lambda<1$ is reported in the SI.

We highlight two key observations. First, the exponent $\lambda$ dictates the abruptness of the yielding process. Second, the decay exponents of the moduli are insensitive to the value of $\lambda$, pointing toward a universal scaling: the first-harmonic moduli exhibit exponents of $-1.5$ and $-1$, in excellent agreement with those observed in experiments and derived from the ERSS approximation. 

A representative full stress evolution for increasing $\gamma_0$ is shown in figure \ref{fig2}-d, alongside the stress at zero strain, $\sigma_0$ (Fig. \ref{fig2}-e), as a function of the strain amplitude. We recover the behavior observed experimentally, with $\sigma_0$ showing linear viscoelastic, superlinear/yielding, and a fluidized regime featuring a constant stress plateau. 
Yielding in particular, can be further linked to $\lambda$.\\ 
We evaluate for this purpose the viscoplastic fragility $\Phi$ for various parameter sets (details in SI). We define $\Phi$ similarly to Ref. \cite{donleyElucidatingOvershootSoft2020}. 
We define $\Phi$ as the maximum slope of the normalized loss modulus increment. Specifically, let $\Delta G'' = G''(\gamma_0) - G''_l$ be the increment over the linear contribution $G''_l = \frac{G_c\omega}{\Gamma_0(1+\omega^2/\Gamma_0^2)}$; the fragility $\Phi$ is given by the maximum derivative of $\Delta G''$ (normalized by its peak value $\Delta G''_M$) with respect to the strain amplitude scaled by the position of the peak $\gamma_M$:
\begin{flalign}
    & \Phi = \max\left\{\frac{\gamma_M}{\Delta G''_M} \frac{d(\Delta G'')}{d\gamma_0}\right\}. &&
\end{flalign}

This quantity is experimentally accessible through simple dynamic strain sweeps, provided the sampling resolution near yielding is sufficient to compute the first derivative with high numerical accuracy.  
The result is shown in figure \ref{fig2}-c.
Quite strikingly, for overdamped systems ($r \gg \omega$), $\Phi$ is found to be governed solely by $\lambda$, which serves as single scaling parameter that describes the transition from ductile to brittle yielding. $\Phi(\lambda)$ diverges at $\lambda = 2$, exhibiting hallmarks of criticality: when plotted against the reduced exponent $\varepsilon = |2 - \lambda|/2$, the fragility follows a power-law scaling $\Phi(\varepsilon) = A \varepsilon^B$, with $A = 1.28 \pm 0.02$ and $B = -0.504 \pm 0.008$.
The divergence of the viscoplastic fragility as $\varepsilon \to 0$ with a critical exponent of $\approx -1/2$ strongly suggests a mean-field scenario, reminiscent of Landau-like behavior \cite{stanleyIntroductionPhaseTransitions2010}, where fluctuations are implicitly averaged out. 
Additionally, we observe that the BFF model predicts a viscoplastic fragility that remains bounded from below even when considering the regime where $\lambda < 1$: we find $\Phi(\lambda)\geq \Phi(\lambda_m)\simeq$ $1.674\pm 0.038$, with $\lambda_m=0.491\pm 0.024$ (SI).

While this is corroborated by experimental data (Figure S7), it warrants a dedicated experimental campaign. 

\subsection*{Entropy production and switching entropy}
\begin{figure}
    \centering
    \makebox[\textwidth][c]{
    \begin{minipage}{1.4\textwidth}
    \centering
    \includegraphics[width=\linewidth]{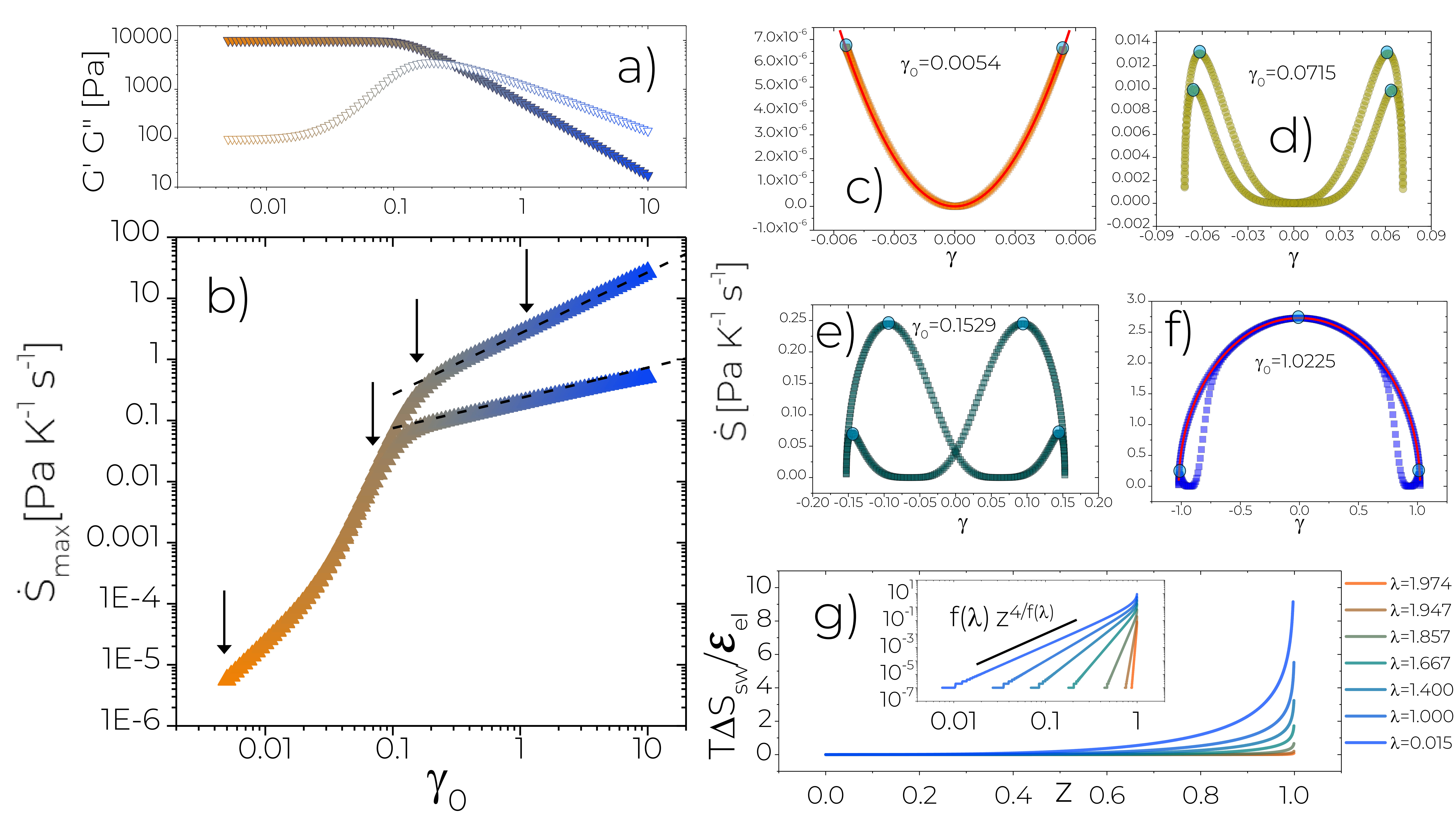}
        \caption{\textbf{Entropy production and switching entropy}. First-harmonic moduli \textbf{a)} and relative entropy production maxima $\dot{S}_{max}$ as a function of the strain amplitude \textbf{b)} obtained for the same parameters as in figure \ref{fig2}-d (same color code). The arrows show those amplitude for which the full entropy production $\dot{S}$ is shown in panels \textbf{c)}, \textbf{d)},\textbf{ e)} and \textbf{f)}. The dashed lines show the entropy production maxima predicted by the approximate equations S29 and S32 (SI). The light-blue circles in panels c), d) e) and f) mark the maxima of $\dot{S}$. The red lines in panels c) and f) show the entropy production predicted by equations S27 and S28 (SI), which are respectively a parabola and a semi-ellipse in the $\dot{S}-\gamma$ plane. Panel \textbf{g)} shows the switching entropy-to-recoverable energy ratio as a function of $z=\sigma/\sigma_p$ computed for varying lambda as indicated in the panel. The inset shows the same curves in log-log scale.}\label{fig3}
    \end{minipage}
    }
\end{figure}

By solving Eqs. (\ref{fluidity1}) and (\ref{fluidity5}) for stress and fluidity, we can push this analysis a step further and determine the rate of energy dissipation per unit volume, $T\dot{S}$, where $T$ is the absolute temperature and $\dot{S}$ the internal and irreversible entropy production rate \cite{hauptContinuumMechanicsTheory2002,glansdorffThermodynamicTheoryStructure1971} (SI):
\begin{flalign} 
&T\dot{S} = \sigma(t)\,\dot{\gamma}(t) - \dot{\mathcal{E}} = \frac{\sigma^2(t)}{G_c}\left[\Fl(t) + \Gamma_0(\omega)\right].&&
\end{flalign}
Here $\sigma(t)\,\dot{\gamma}(t)$ represents the injected power per unit volume and $\mathcal{E}(t) = \sigma^2(t) / (2G_c)$ accounts for the stored elastic energy density. We thus resolve the intra-cycle energy dissipation as a function of the strain amplitude $\gamma_0$. Figure \ref{fig3}-a,b shows a dynamic strain sweep (same parameters as in fig.\ref{fig2}-d) and the corresponding intra-cycle maxima of the entropy production $\dot{S}_{max}$, where we observe the emergence of two maxima at yielding.  

To elucidate this phenomenon in figure \ref{fig3}-c,d,e,f we plot $\dot{S}$ for four representative amplitudes. In the linear viscoelastic regime, the system dissipates the greatest amount of energy at the flow reversal points, where the viscous strain rate — and the stress — reaches its maximum. In this regime, $\dot{S}(\gamma)$ follows a quadratic form (SI).


Beyond the linear regime, entropy production is no longer a simple function of strain $\gamma(t)$. Instead, fluidization and plasticity generate multiple maxima. In the fluidized state, $\dot{S}(\gamma)$ converges to a characteristic form invariant across amplitudes beyond yielding (Fig. \ref{fig3}-f). This state features a primary maximum $\dot{S}_{max}^{(1)}$ at the peak shear rate, and secondary maxima $\dot{S}_{max}^{(2)}$ driven by energy dissipation during stress inversion between steady-stress states. 

The ratio $\dot{S}_0/\dot{S}_{max}^{(1)}$, where $\dot{S}_0$ is the entropy production rate at zero strain, transitions from 0 to 1 across yielding and serves as a thermodynamic marker of fluidization, bypassing traditional alternative analyses based on harmonic \cite{hyunReviewNonlinearOscillatory2011} or polynomial \cite{ewoldtNewMeasuresCharacterizing2008} stress decompositions.

Figure \ref{fig3}-b tracks the evolution of the entropy production maxima throughout the dynamic strain sweep. Interestingly, the two maxima follow distinct power-laws: the peak associated with the maximum shear rate scales linearly with $\gamma_0$, whereas the maximum associated with stress inversion exhibits a lower power-law exponent, and can be well approximated by Eq. (S32) (SI), predicting a scaling $\sim\gamma_0^{1/2}$. 

This brings us to quantify the \textit{switching entropy} — the entropy variation produced during the transition between steady stresses. More generally we compute the switching entropy between $-\sigma$ and $+\sigma$ as $\Delta S_{sw}(\sigma)$=$\int_{-\sigma}^{\sigma} \dot{S}(\tilde{\sigma})\frac{d\tilde{\sigma}}{\dot{\tilde{\sigma}}}$, which admits a closed-form expression for large amplitudes ($\Gamma_0/(\gamma_0\omega)\to 0$) and reads (SI):
\begin{flalign}\label{fluidity7}
\Delta S_{sw}(z)=\frac{1}{T}f(\lambda)\,\mathcal{E}_{el}\,z^{\frac{4}{f(\lambda)}}\,{}_{2}F_{1}\left[1,\frac{1}{2-f(\lambda)},\frac{3-f(\lambda)}{2-f(\lambda)},z^{4\left(\frac{2}{f(\lambda)}-1\right)}\right],
\end{flalign}
where $f(\lambda)=\frac{2(2-\lambda)}{3-\lambda}$, $z=\frac{\sigma}{\sigma_p}$, ${}_{2}F_{1}(a,b,c,z)$ the Gaussian hypergeometric function, and $\mathcal{E}_{el}=\sigma_p^2/(2G_c)$ is the energy stored in the steady-stress state. The material function $T\Delta S_{sw}(z)/\mathcal{E}_{el}$ is plotted in figure \ref{fig3}-g.
The pre-factor $\frac{1}{T}f(\lambda)\,\mathcal{E}_{el}$ represents the switching entropy that would be generated by considering exclusively the elastic recoil, $\dot{\sigma} \simeq \dot{\sigma}_r = G_c \dot{\gamma}$, i.e. in the absence of the feedback mechanism between fluidity and stress. In this case the time needed to the complete switch is finite and reads $\tau_d=\sqrt{\frac{4\sigma_p}{G\omega^2\gamma_0}}$ (SI). Under these conditions, the switching entropy from $-\sigma_p$ to $+\sigma_p$ converges to a fraction of the recoverable energy dictated by the system fragility:
\begin{equation}\label{fluidity8}
\frac{T\Delta S_{sw}^{(el)}}{\mathcal{E}_{el}}=f(\lambda),
\end{equation}
which brings to the following expression for the stationary (dynamic yield) stress $\sigma_p$

\begin{equation}\label{fluidity9}
\sigma_p=\sqrt{\frac{2G_cT\Delta S_{sw}^{(el)}}{f(\lambda)}}
\end{equation}

Equations \eqref{fluidity8} and \eqref{fluidity9} establish the coupling between viscoplastic fragility, recoverable energy, and the dissipation produced when transitioning between counter-flowing steady states, thereby inextricably linking the value of the dynamic yield stress to symmetry breaking and to the underlying thermodynamics governing the steady-state switch. 

As already shown by strain sweeps, $\lambda$ dictates the specific energy dissipation pathway during a steady-state switch. As $\lambda \to 2$, the system dissipates progressively less energy for small inversions around zero stress, following a power law of the form $f(\lambda)\,z^{4/f(\lambda)}$. A logarithmic divergence emerges as $z \to 1$ (SI), reflecting the infinite time required to switch between states when the residence-time measure in stress space $d\sigma/\dot{\sigma}$ diverges, while the ratio $\frac{T\Delta S_{sw}^{(el)}}{\mathcal{E}_{el}}$ can be graphically determined by extrapolating the power laws to $z=1$ (inset of figure \ref{fig3}-g). 
$f(\lambda)$ thus defines the coupling coefficient, bridging steady elastic energy, switching entropy and viscoplastic fragility; it couples the recoverable energy to the minimum structural entropy the system must "pay" to transition from one steady state to the other. Remarkably, we also observe that the unrecoverable plastic component of the first-harmonic loss modulus ($G''_p$) \cite{donleyElucidatingOvershootSoft2020} rises in the linear regime following a power law, $G''_p\sim\gamma_0^{\frac{2}{2-\lambda}}=\gamma_0^{\frac{2[2-f(\lambda)]}{f(\lambda)}}$ (SI). 
The interplay between the steady-state recoverable energy and the switching entropy thus dictates the non-linear response of yield-stress fluids at any amplitude.

\begin{figure}[htbp]
    \centering
    \makebox[\textwidth][c]{
    \begin{minipage}{1.4\textwidth}
    \centering
    \includegraphics[width=\linewidth]{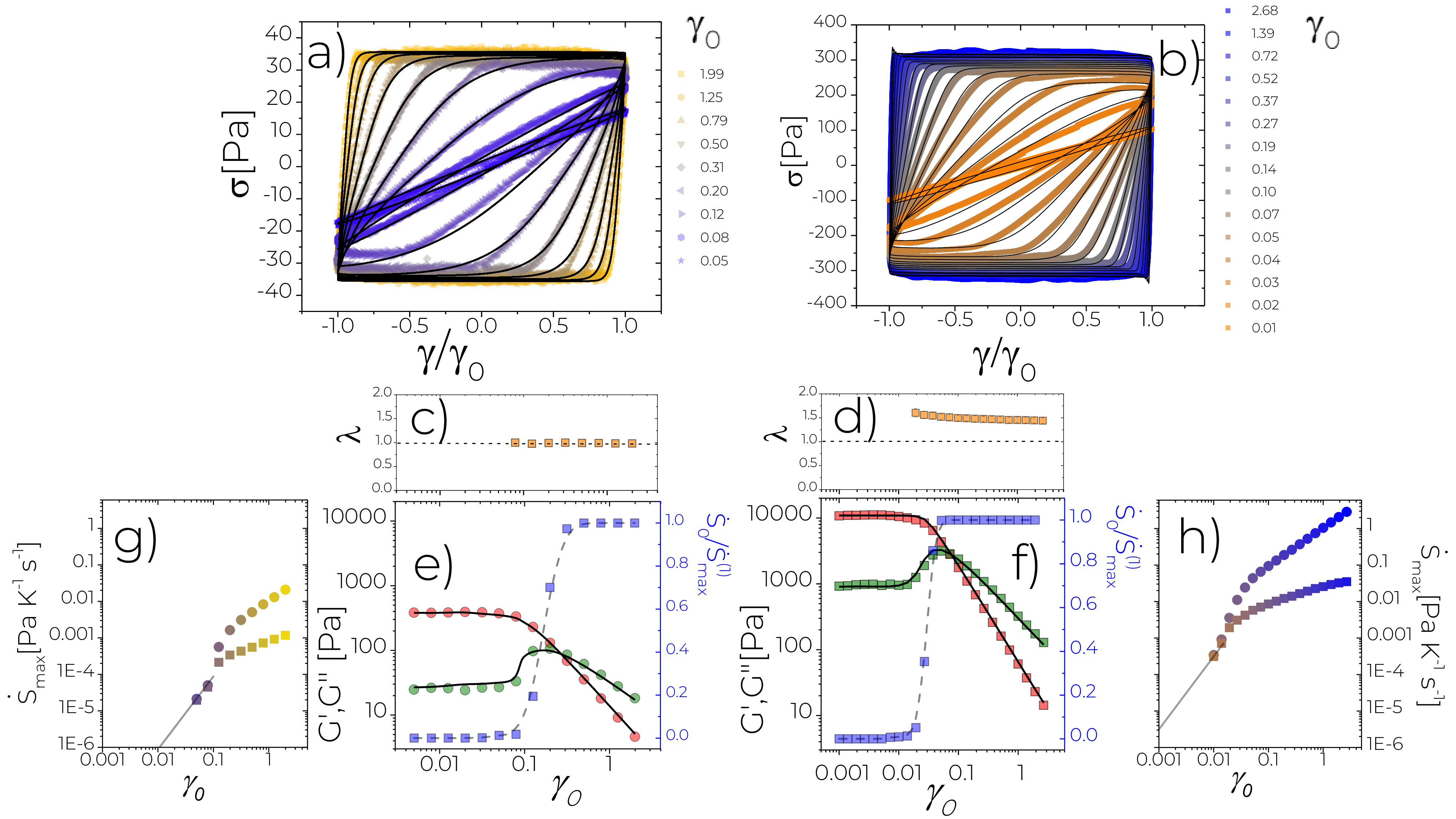}
    \caption{\textbf{Comparison between theory and experiments}: Full intracycle stress response $\sigma(\gamma/\gamma_0)$ for jammed microgels MCr5 ($\varphi$=1.3) a) and glassy Ludox TM50 suspensions ($\varphi$=0.432) b). Solid lines are the best fit of the BFF model to the data. The resulting $\lambda$-values in the non-linear regime are shown in panels c) and d). Panels e) f) show the comparison between the first harmonic moduli inferred from BFF fit to the data (solid lines) and the moduli measured by strain-controlled amplitude sweeps (filled points). The entropy production rate ratio $\dot{S}_0/\dot{S}_{max}^{(1)}$ obtained from the best fit is also reported (blue squares - right axis). Panels g) and h) show the entropy production maxima obtained from the fits shown in panel a) and b) for jammed microgels and glassy Ludox, respectively. The solid gray lines correspond to the unique maximum obtained in the linear regime according to equation S27: $\dot{S}_{max}=\frac{G_c\Gamma_0}{T}\gamma_0^2$.}\label{fig4}
    \end{minipage}
    }
\end{figure}

To assess the predictive capability of the BFF, we performed a direct comparison with Large Amplitude Oscillatory Shear (LAOS) experimental data for the two yield-stress systems whose intra-cycle stress response has been measured (Fig.~\ref{fig4}-a,b). 

The BFF model was employed to fit the full stress-strain cycles $\sigma(\gamma)$ across the entire range of investigated strain amplitudes, spanning the linear regime, the yielding transition, and the fluidized state. 
To ensure physical consistency, $G_c$ was fixed at each amplitude to the value of the apparent cage modulus, defined as $G_c = \left. \frac{d\sigma}{d\gamma} \right|_{\sigma=0}$ \cite{rogersSequencePhysicalProcesses2011,truzzolilloRheologicalDetectionCaging2013a}, which is nearly amplitude-independent (SI). At the microscopic level, $G_c$ describes the weak, recoverable elastic stretching of the pristine microstructure right before the onset of yielding and flow under large deformations \cite{rogersSequencePhysicalProcesses2011}.

$\lambda$ and $\sigma_c$ were the only free fitting parameters, which is equivalent to leaving free $\Delta S_{sw}^{(el)}$ and the recoverable energy $\mathcal{E}_{el}$. We observe that the model's output is largely insensitive to the precise value of the damping factor $r$, provided the condition $r \gg \omega$ is satisfied, consistent with our overdamped assumption. We obtained an excellent agreement between the BFF model and experimental data for both systems, as demonstrated by the accurate capture of both the full stress-strain Lissajous-Bowditch diagrams (Fig.~\ref{fig4}-a,b) and the first-harmonic moduli (Fig.~\ref{fig4}-e,f).

Remarkably, the fitting procedure yields distinct values of $\lambda$ (Fig.~\ref{fig4}-c,d) for the two systems. These values were successfully extracted within the non-linear regime, and point to a larger brittleness of the Ludox suspension compared to the jammed microgels ($\lambda_{\text{Ludox}} > \lambda_{\text{MCr5}}$), in agreement with a previous study \cite{aimeUnifiedStateDiagram2023a}. At the maximum strain amplitude investigated in this work, glassy Ludox and jammed microgels are thus characterized by a ratio $f(\lambda)$ equal to $0.72$ and $1.00$ respectively. 

In both cases the entropy production ratio $\dot{S}_0/\dot{S}_{max}^{(1)}$ varies continuously from $0$ to $1$ crossing yielding (Fig.~\ref{fig4}-e,f), with a sharper transition (linear scale in SI) and larger maxima of entropy production rates (Fig.~\ref{fig4}-g,h) observed for the Ludox suspension.

We observe also that the $\lambda$-values for the Ludox suspension, ranging between 1.60 and 1.44, show a subtle, yet discernible, dependence on the strain amplitude. We attribute this trend mostly to the progressive rejuvenation of the system as the forcing increases. Ludox \cite{truzzolilloBulkInterfacialStresses2015c} are considerably smaller than microgels \cite{philippeGlassTransitionSoft2018c,elancheliyanImpactPolyelectrolyteAdsorption2023a}, they drift toward equilibrium significantly faster, are better annealed and show reduced ductility \cite{ozawaRandomCriticalPoint2018,divouxDuctiletobrittleTransitionYielding2024}. 
We finally note that the jammed microgels exhibits a value of $\lambda$ close to unity, which represents the threshold for the $C^1$ continuity of the BFF functional $\mathcal{V}$. While it is established that the limit $\lambda \to 2$ marks the ductile-to-brittle transition 
the physical significance of the $\lambda < 1$ threshold remains an open question,
which is mostly motivated by the fact that all values of $\lambda$ extracted from existing viscoplastic fragility data, are compatible with $\lambda \geq 1$ (SI) and that $\lambda<1$ gives rise to a counterintuitive increase of the constant $\alpha$ for decreasing $G_c$ (SI). A dedicated experimental campaign is in order to accurately measure fragility and unravel this point.\\
\section*{Discussion}
The BFF developed here establishes a unifying paradigm for the rheology of yield-stress materials under oscillatory shear, with a phenomenology emerging from universal elastoviscoplastic scaling laws. By identifying fluidity as a fast variable, we have transcended the limitations of classical constitutive descriptions, revealing the NESS architecture of yield-stress fluids where the steady-state stress is dictated by the symmetry breaking of the system.

Central to this framework is the derivation of an unprecedented thermomechanical identity for the dynamic yield stress: equations \eqref{fluidity8} and \eqref{fluidity9} 
establish the existence of a stress threshold as the direct consequence of the interplay between the  energy stored in the plastic state and a finite switching entropy magnitude $\Delta S_{sw}^{(el)}$, which are coupled by viscoplastic fragility. 

The success of our ''frozen-time'' Lyapunov function in describing experimental data suggests that a unique scalar exponent, $\lambda$, is the fundamental metric governing the yielding transition. This parameter does not merely describe plasticity production; it encodes the abruptness of yielding—from ductile to brittle—by governing the rate of entropy production during the cycle. 

Beyond theoretical implications, our BFF provides a robust novel analytical tool for the rational physical characterization and design of yield-stress materials.

\subsection*{Limitations and Perspectives}
While the proposed Bistable Fluidity Framework (BFF) successfully captures the macroscopic non-linear response and universal scaling behaviors of yield-stress fluids, it inherits a clear mean-field character and does not directly address the microscopic origin of plasticity generation. From this perspective, a natural and promising extension of the current framework would be the inclusion of spatial gradient terms in the governing function $\mathcal{V(\theta)}$, thereby accounting for the stress spatial heterogeneities and local fluctuations in the instantaneous relaxation rate characteristic of soft glassy systems. This represents both a main limitation of the present work and a compelling pathway for future extensions, which is currently under active investigation. 
An additional limitation—which nevertheless does not affect the quantitative robust conclusions drawn here—stems from the current lack of high-resolution strain-amplitude experimental datasets in the immediate vicinity of the yielding transition during dynamic strain sweeps. Acquiring such finely resolved data will be crucial to compute the first derivatives with sufficient numerical accuracy and, consequently, to evaluate the viscoplastic fragility $\Phi$ unambiguously across different complex fluids.

\backmatter

\section*{Methods}

\subsection*{Sample synthesis and preparation}
The synthesis of MCr5 microgels (with 5.4 mol\% of crosslinker) performed via emulsion polymerization and the preparation of their suspensions are described in detail in Ref.~\cite{elancheliyanImpactPolyelectrolyteAdsorption2023a}. The synthesis and sample preparation protocols for the MCr1 microgels (with 1.4 mol\% of crosslinker) are reported in Refs.~\cite{senffTemperatureSensitiveMicrogel1999,philippeGlassTransitionSoft2018c}. Ludox TM50 particles were purchased from Sigma-Aldrich and used as received without further purification. Glassy ludox suspensions were prepared by centrifugation as in Refs. \cite{philippeGlassTransitionSoft2018c,truzzolilloBulkInterfacialStresses2015c}. The effective volume fraction $\varphi$ of the microgel suspensions was determined by matching the zero-shear viscosity of dilute samples with the Einstein-Batchelor equation, as comprehensively detailed in Refs.~\cite{elancheliyanImpactPolyelectrolyteAdsorption2023a,philippeGlassTransitionSoft2018c}. Conversely, the volume fraction of the Ludox suspensions was determined by measuring the dry weight residual of each sample alongside the known mass density of the silica particles \cite{truzzolilloBulkInterfacialStresses2015c}.
\subsection*{Rheological characterization and LAOS signal extraction}
\subsubsection*{Nonlinear rheology tests}

All rheological measurements were performed using a strain-controlled ARES RFS 1KFRT rheometer (TA Instruments) equipped with a torque transducer with a working range of 0.002 g·cm to 20 g·cm. For all the tests we employed a cone-plate geometry (cone angle $1^\circ$ and diameter $25$ mm). All measurements were carried out at a temperature of $20.00^\circ$C $\pm$ $0.01 ^\circ$C controlled via a Peltier element. To prevent evaporation during measurements, the sample gap was sealed with silicone oil. Prior to each measurement, a pre-shear was applied for 180 s at a strain amplitude of $250\%$ and a frequency of $\omega=1$ rad/s, in order to erase any loading history.

Dynamic strain sweep (DSS) measurements were performed at a fixed frequency, with the strain amplitude $\gamma_0$ spanning 3 order of magnitudes. This allowed a comprehensive characterization the initial solid-like response (linear regime), the yielding transition and the fully fluidized state.
Flow curves were obtained by imposing descending ramps of steady shear rates from $300\text{ s}^{-1}$ to $0.01\text{ s}^{-1}$ for PNIPAM systems and from $100\text{ s}^{-1}$ to $0.01\text{ s}^{-1}$ for Ludox suspensions, holding the shear rate constant for $30\text{ s}$ at each point along the curve. Prior to each ramp, step-rate experiments were performed at selected shear rates to evaluate the time required for the samples to reach a stable steady-state stress, which was found to be shorter than $30\text{ s}$. 

\subsubsection*{Stress signal acquisition}
The acquisition of the raw torque signal has been obtained through a National Instruments NI USB-6008 acquisition board. The system is controlled by software developed in Python 3.0, utilizing the NI-DAQmx driver software for National Instruments boards.
Using an integrated system directly connected to the ARES RFS rheometer is not possible. Indeed, the low input impedance of the acquisition board affects the rheometer's electrical signals, which distorts the results.  It was necessary to build interface electronics between the rheometer and the acquisition board to achieve an input impedance greater than 500 M$\Omega$. The circuit is based on a Texas Instruments TL074 chip : FET-input operational amplifier 
The python code developed features functions for data exchange with the acquisition board, signal display, and data recording [Add public link upon publication]. The experimental sampling interval and 
the number of cycle per amplitude were set to $\Delta t_s = 0.1$\,s and $N_{cycle}=7$ respectively. To ensure sufficient numerical accuracy of the fitting procedure based on the integration of the governing ordinary differential equations (ODEs) (detailed below) and to guarantee a faster convergence of the fitting algorithm, the data were subsequently interpolated using a cubic spline function. 
\subsection*{Numerical integration of the BFF model}
The coupled differential equations \eqref{fluidity1} and \eqref{fluidity5} governing the evolution of stress $\sigma(t)$ and fluidity $\Fl(t)$ were solved numerically using a first-order forward Euler finite-difference scheme. To simulate the material response under Large-Amplitude Oscillatory Shear (LAOS), we prescribed a sinusoidal strain input $\gamma(t) = \gamma_0 \sin(\omega t)$. The system was discretized with a constant time step $\Delta t = 5 \times 10^{-5}$~s. This value was chosen to be several orders of magnitude smaller than both the period of oscillation $T = 2\pi/\omega$ and the characteristic relaxation time of the system, ensuring numerical stability and convergence across the explored range of strain amplitudes. 

At each time step $i$, the state variables were updated according to the following discretized forms:
\begin{equation}
\sigma(t_{i+1}) = \sigma(t_i) + \left\{ G_c \dot{\gamma}(t_i) - \sigma(t_i) \left[\Gamma_0 + \Fl(t_i)\right] \right\} \Delta t
\end{equation}
\begin{equation}
\Fl(t_{i+1}) = \Fl(t_i) + \left[ -r \Fl(t_i) \left( 1 - \left| \frac{\sigma(t_i)}{\sigma_c} \right|^\lambda \left| \frac{\dot{\gamma}(t_i)}{\Fl(t_i)} \right|^{2-\lambda} \right) \right] \Delta t
\end{equation}
To ensure that the system reached a non-linear oscillatory steady state (NOSS) and to eliminate initial transients, each simulation was run for $N_{cycle} = 100$ cycles. 

The first-harmonic viscoelastic moduli, $G'$ and $G''$, were extracted directly from equations S52 and S53 (SI). The model was integrated for strain amplitudes ($\gamma_0$) ranging from $10^{-3}$ to $10$, allowing for a comprehensive characterization of the yielding transition. All numerical procedures were implemented in a custom C script.

\subsection*{Fitting procedure of the intra-cycle stress response}

For each strain amplitude, the plasticity production exponent $\lambda$ and $\sigma_c$ were determined by fitting the model to the experimental stress time series $\sigma(t)$. All other parameters ($G_c, \Gamma_0, r$) were fixed prior to the fitting procedure. $G_c$ was measured for each amplitude. $\Gamma_0$ is set by the ratio between the viscoelastic moduli in the linear regime : $\Gamma_0=\omega \frac{G''_l}{G'_l}$. $r$ was fixed to $150$ s$^{-1}$, which is much larger than the  angular frequency $\omega$ set in the experiment. The noise signal hampers to determine the exact value of $r$ in the overdamped regime, where its value does not affect considerably the sum of residuals. 

The numerical integration scheme follows that described in the preceding section, with the time step $\Delta t$ depending on the samples, $\Delta t=0.006$ s for MCr5 sample and $0.0006$ s for the Ludox sample. 
To ensure that the simulated stress response had reached a steady periodic state, we simulated 16 stress cycles per amplitude for MCr5 sample and 40 cycles the Ludox sample for each set of fitting parameters. Only the final cycle of the simulation, was retained for the fitting scheme.
The fitting procedure was performed iteratively using a Levenberg-Marquardt least-squares algorithm [reference to add upon publication], involving two free parameters: $\sigma_c$ and $\lambda$. Optimization is carried out sequentially as follows: at the largest strain amplitude, we initiate the algorithm with an initial guess for $\lambda$ within the physically admissible range ($0 < \lambda \le 2$) to determine the best-fit value of $\sigma_c$. This value of $\sigma_c$ is subsequently fixed, and the model is refitted by leaving $\lambda$ as the sole free parameter. This two-step process is repeated iteratively until full convergence of both parameters is achieved. The final optimized parameters obtained at the highest amplitude are then utilized as the initial seeds for the fitting procedure at progressively lower strain amplitudes.
The model consistently achieved $R^2>0.97$ across all strain amplitudes. The procedure was applied independently for each strain amplitude. 

\bmhead{Supplementary information}

The online version of this article contains supplementary material, which includes a detailed overview of the non-linear rheology of jammed microgels and Ludox suspensions (flow curves and strain sweeps) along with their cage moduli; the analytical derivation of the first-harmonic moduli based on the elastic-recoil plus steady-stress (ERSS) approximation; the framework of the elastoviscoplastic Maxwell and BFF models, including tracking solutions, bifurcation diagrams, and the physical significance of the model parameters; an extensive evaluation of viscoplastic fragility for varying model parameters; and a comprehensive analysis of entropy production rates, switching entropies.




\bmhead{Acknowledgements}
D.T. warmly thanks Luca Cipelletti, Laurence Ramos, Stefano Aime and Dimitris Vlassopoulos for insightful discussions. This work was supported by the Agence Nationale de la Recherche under grant number [ANR-20-CE06-0030-01; THELECTRA].

\bmhead{Competing interest}
The authors declare no competing interests.

\bmhead{Author contribution}
D.T developed the BFF model and designed experiments. R.E. and D.T. analysed data. J.M.F. configured the setup to extract the stress from strain-controlled experiments and developed the acquisition software. E.C. synthesized PNIPAM microgels. D.T. R.E. and J.M.F. wrote the manuscript. D.T. supervised the project. All authors contributed to the discussion of the results and revision of the manuscript.









\begin{appendices}






\end{appendices}

\bibliography{Universal_Scaling_LAOS}

\end{document}


\title[Supplementary Information:\\ Universal scalings and switching entropy in yield-stress fluids]{Supplementary Information: Universal scalings and switching entropy in yield-stress fluids}


\author[1]{\fnm{Rajam} \sur{Elancheliyan}}\email{rajam.elancheliyan@gmail.com}

\author[1]{\fnm{Jean Marc} \sur{Fromental}}\email{jean-marc.fromental@umontpellier.fr}

\author[1]{\fnm{Edouard} \sur{Chauveau}}\email{edouard.chauveau@umontpellier.fr}

\author*[1]{\fnm{Domenico} \sur{Truzzolillo}}\email{domenico.truzzolillo@umontpellier.fr}

\affil*[1]{\orgdiv{L2C}, \orgname{Univ Montpellier, CNRS, F-34095 Montpellier, France}, \orgaddress{\city{Montpellier}, \postcode{F-34095}, \country{France}}}


\maketitle

\clearpage
\singlespacing 
\tableofcontents
\clearpage
\onehalfspacing 

\section{Non-linear rheology of jammed microgels and Ludox suspensions}
\begin{figure}[htbp]
    \centering
    \includegraphics[width=0.9\linewidth]{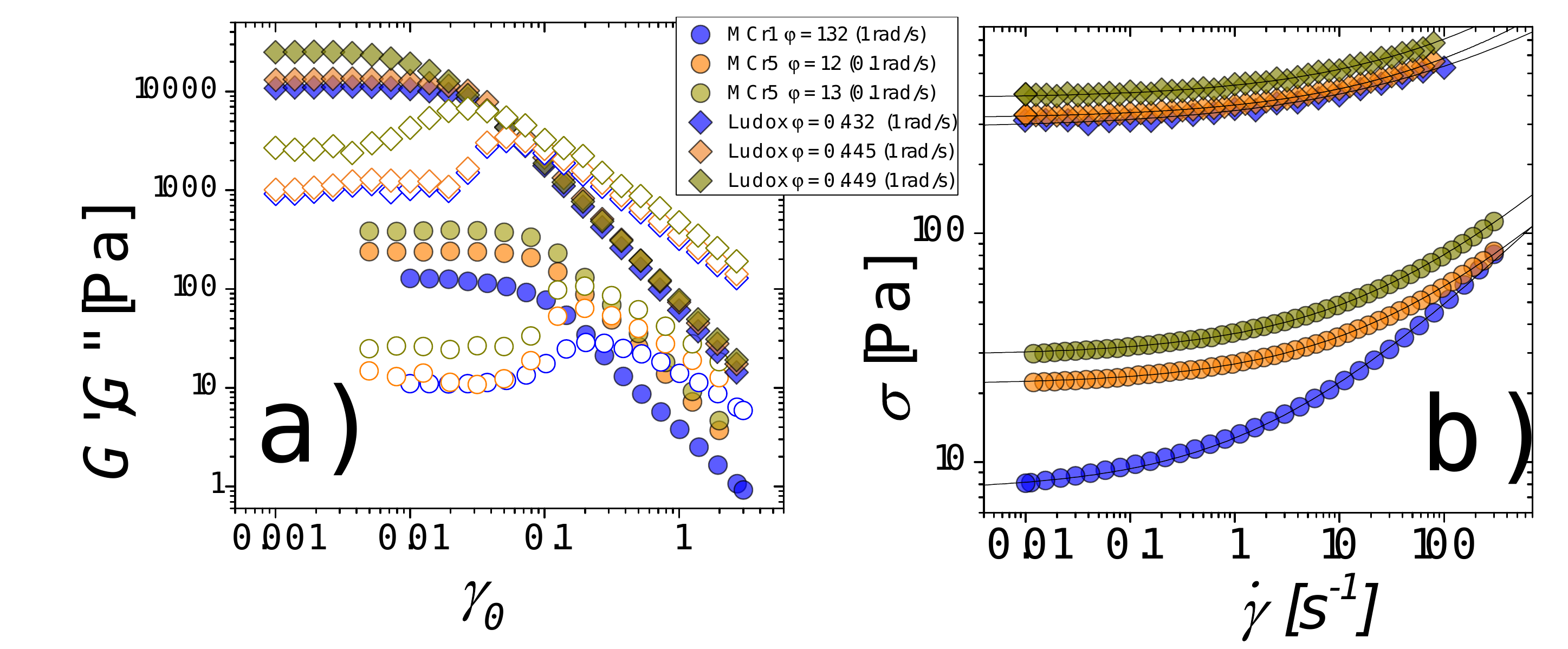}
    \caption{\textbf{Non-linear rheological characterization of microgel and Ludox suspensions.} 
    \textbf{(a)} Dynamic strain sweeps showing the storage modulus $G'$ (filled symbols) and loss modulus $G''$ (open symbols) as a function of strain amplitude $\gamma_0$. The investigated samples encompass two distinct model systems—jammed microgels (MCr1 and MCr5) and colloidal silica (Ludox) particles at various volume fractions $\varphi$—effectively spanning approximately two orders of magnitude in the linear storage modulus. The angular frequency of the sweeps is indicated in the legend.  
    \textbf{(b)} Corresponding steady-state flow curves  $\sigma(\dot{\gamma})$ (same symbols). The macroscopic dynamic yield stress exhibits a comparable two-order-of-magnitude variations in the dynamic yield stress across the different samples. Solid lines represent the best-fit regressions to the experimental data using the Herschel-Bulkley constitutive model ($\sigma = \sigma_Y + K\dot{\gamma}^n$). The extracted best-fit parameters ($\sigma_Y$, $K$, and $n$) are comprehensively summarized in Table~\ref{tab:herschel_bulkley_parameters}.}
    \label{fig:rheoextended}
\end{figure}

\begin{table}[htbp]
\centering
\caption{\textbf{Herschel-Bulkley fitting parameters for microgel suspensions and colloidal silica particles.} The constitutive equation $\sigma = \sigma_Y + K\dot{\gamma}^n$ was utilized to fit the steady-state flow curves displayed in Fig.~\ref{fig:rheoextended}(b). The table summarizes the optimal values for the dynamic yield stress $\sigma_Y$, the consistency index $K$, and the exponent $n$ along with their respective experimental uncertainties.}
\label{tab:herschel_bulkley_parameters}
\vspace{0.2cm}
\begin{tabular}{l c c c}
\toprule
\textbf{System / Sample} & $\sigma_Y$ (Pa) & $K$ (Pa$\cdot$s$^n$) & $n$ (--) \\
\midrule
MCr1 ($\varphi = 1.32$)  & 7.5 $\pm$ 0.5 & 5.24 $\pm$ 0.08 & 0.451 $\pm$ 0.003 \\
MCr5 ($\varphi = 1.2$)   & 21.84 $\pm$ 0.08 & 4.99 $\pm$ 0.07 & 0.433 $\pm$ 0.003 \\
MCr5 ($\varphi = 1.3$)   & 29.3 $\pm$ 0.1 & 7.2 $\pm$ 0.1 & 0.427 $\pm$ 0.003 \\
Ludox ($\varphi = 0.432$) & 288 $\pm$ 5 & 56 $\pm$ 6 & 0.32 $\pm$ 0.02 \\
Ludox ($\varphi = 0.445$) & 317 $\pm$ 1 & 46 $\pm$ 1 & 0.379 $\pm$ 0.005 \\
Ludox ($\varphi = 0.449$) & 389 $\pm$ 3 & 54 $\pm$ 3 & 0.38 $\pm$ 0.01 \\
\bottomrule
\end{tabular}
\end{table}

\newpage

\section{Cage modulus of MCr5 ($\varphi = 1.3$) and Ludox ($\varphi = 0.432$) suspensions}

\begin{figure}[htbp]
    \centering
    \includegraphics[width=0.75\linewidth]{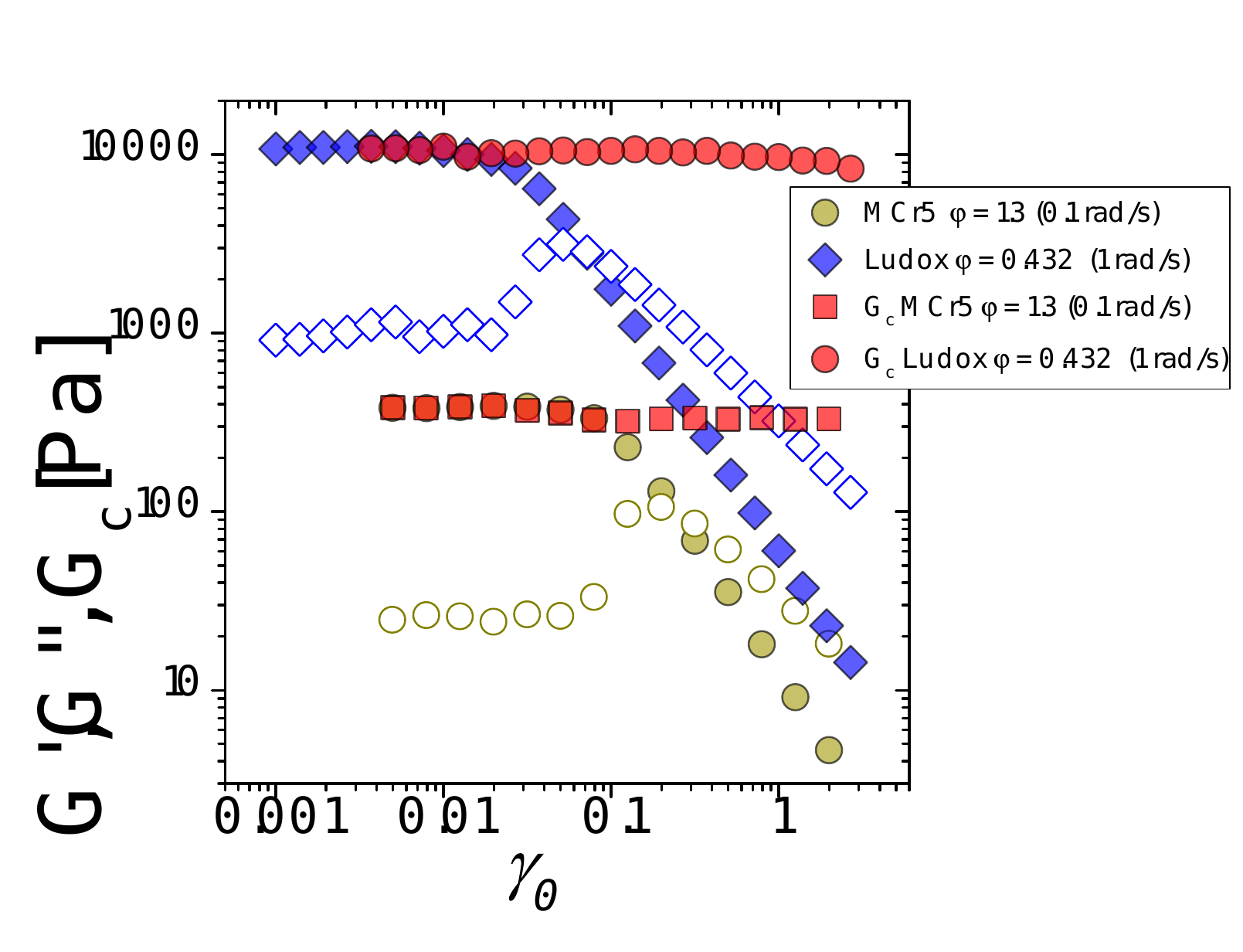}
    \caption{\textbf{Strain-amplitude dependence of the cage modulus.} 
    The dynamic cage modulus $G_c$, evaluated directly from the intra-cycle elastic response as $G_c = \left. d\sigma / d\gamma \right|_{\sigma = 0}$ (red symbols), is plotted as a function of the strain amplitude $\gamma_0$. 
    Profiles are shown for the two yield stress systems whose detailed intra-cycle response is discussed of the main text (Fig. 4): the MCr5 microgel suspension at an effective volume fraction of $\varphi = 1.3$ probed at $\omega = 0.1$\,rad/s and the Ludox TM 50 colloidal suspension at a volume fraction of $\varphi = 0.432$ probed at $\omega = 1$\,rad/s. 
    The corresponding first harmonic moduli $G'$ and loss $G''$ moduli profiles are reported for comparison.}
    \label{fig:cage_modulus_evolution}
\end{figure}

\section{Universal decay of the first-harmonic moduli from ERSS approximation of the stress wave form}

To derive the asymptotic scaling of the viscoelastic moduli, we expand the stress evolution around the flow reversal point where $\dot{\gamma}=0$. We define the stress in the elastic restoring region, from $-\sigma_p$ to +$\sigma_p$, as $\sigma_r(t)$. In the vicinity of the flow reversal ($\dot{\gamma}=0$) we can adopt the following elastic-recoil approximation:
\begin{align}
    \dot{\gamma} &\simeq \omega^2 \gamma_0 \tau \label{approxgammadot}\\
    \dot{\sigma}_r &\simeq G_c \dot{\gamma} \\
    \sigma_r(\tau) &\simeq \frac{1}{2} \omega^2 G_c \gamma_0 \tau^2 - \sigma_p \label{approxstress}
\end{align}
where $\tau = (t - t_0)$ represents the time elapsed since the flow reversal at $t_0$. 

This approximation holds as long as the condition $G_c\dot{\gamma} \gg \sigma(\Gamma_0 + \Fl(t))$ is satisfied. This is typically the case for systems deeply quenched into the glassy or jammed phase, characterized by vanishingly small (total) relaxation rates at the flow reversal point.

From Equation~\ref{approxstress}, we can calculate the time interval $\tau_d$ required for the stress to transition from $-\sigma_p$ to $+\sigma_p$:
\begin{equation}\label{switchingtime}
    \tau_d = \sqrt{\frac{4\sigma_p}{\omega^2 \gamma_0 G_c}}
\end{equation}

The first-harmonic moduli are then computed by considering the contributions from the two restoring regions and the two constant-stress plateaus ($\sigma=\sigma_p$) occurring within a single period. The Fourier integration yields:
\begin{align}
    G' &\simeq \frac{2\omega}{\pi\gamma_0} \left[ \int_0^{\tau_d} \sigma_r(\tau) \cdot (-1) d\tau + \int_{\tau_d}^{\pi/\omega} \sigma_p \sin(\omega\tau) d\tau \right] \nonumber \\
    &= \frac{16}{3\pi} \frac{\sigma_p^{3/2}}{G_c^{1/2}} \frac{1}{\gamma_0^{3/2}} + O\left(\frac{1}{\gamma_0^{5/2}}\right) \\
    G'' &\simeq \frac{2\omega}{\pi\gamma_0} \int_{\tau_d}^{\pi/\omega} \sigma_p \cos(\omega\tau) d\tau = \frac{4}{\pi} \frac{\sigma_p}{\gamma_0} + O\left(\frac{1}{\gamma_0^2}\right)
\end{align}
In the expression for $G'$, we have utilized the approximation $\sin(-\frac{\pi}{2} + \omega\tau) \simeq -1$ for $\tau \simeq 0$.

By equating the leading terms of both moduli, we determine the characteristic strain $\gamma_y$ at which the two asymptotic laws intersect:
\begin{equation}
    \gamma_y = \frac{16}{9} \frac{\sigma_p}{G_c}    
\end{equation}
Consequently, we can define universal master curves for the fluidized regime as:
\begin{equation}
    \frac{G'}{G_c} = \frac{9}{4\pi} \frac{1}{x^{3/2}} \quad \text{and} \quad \frac{G''}{G_c} = \frac{9}{4\pi} \frac{1}{x}    
\end{equation}
where $x = \gamma_0/\gamma_y$ is the dimensionless strain amplitude. Experimentally, this ubiquitous scaling is recovered by normalizing the moduli by the linear plateau modulus $G_c \simeq G'_{l}$, and scaling the strain amplitude by the value corresponding to the intersection of the two asymptotic power laws.

\newpage

\section{Elastoviscoplastic Maxwell model}
\begin{figure}
    \centering
    \includegraphics[width=1.0\linewidth]{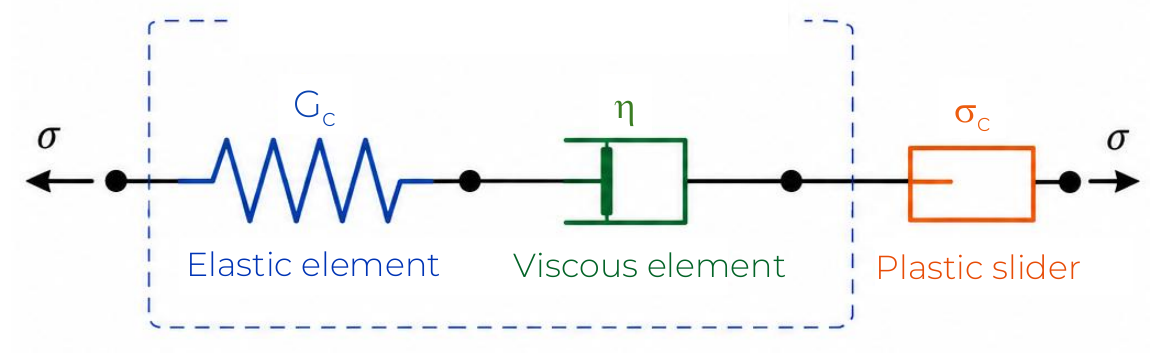}
    \caption{Elastoplastic Maxwell schematic representation of the fluidity framework. The dashed line encloses the two elements of the standard Maxwell model.}
    \label{fig:SM1}
\end{figure}

The dynamical evolution of the stress $\sigma(t)$ in our framework is governed by the equation:
\begin{equation}\label{maxwell}
    \dot{\sigma} = G_c \dot{\gamma} - \left[ \Gamma_0(\omega) + \Fl(t) \right] \sigma
\end{equation}
where $G_c$ is the cage shear modulus and $\Fl(t)$ quantifies the instantaneous relaxation rate arising from plastic rearrangements. This formulation can be mapped onto the elastoviscoplastic mechanical analog shown in Fig.~\ref{fig:SM1}. In this representation, the system is modeled as a series of three distinct elements: an elastic spring with modulus $G_c$, a viscous dashpot with background relaxation $\Gamma_0(\omega)$, and a non-linear plastic slider. 

Under this configuration, the total strain $\gamma$ and the total strain rate $\dot{\gamma}$ are partitioned into elastic, viscous, and plastic contributions:
\begin{equation}\label{gdotds}
    \gamma = \gamma_e + \gamma_v + \gamma_p, \quad \text{and} \quad \dot{\gamma} = \dot{\gamma}_e + \dot{\gamma}_v + \dot{\gamma}_p
\end{equation}
where $\gamma_e = \sigma / G_c$ represents the recoverable elastic strain and $\dot{\gamma}_v = \sigma / \eta(\omega)$ accounts for the linear viscous dissipation at frequency $\omega$. The plastic component, $\dot{\gamma}_p$, is governed by the fluidity $\Fl(t)$, effectively acting as a dissipative "valve" whose activation depends on both the applied stress and the deformation rate. Specifically, the characteristic stress $\sigma_c$ (see main text) regulates the yielding transition, where the sharpness of the plastic onset is dictated by the plasticity production exponent $\lambda$. 
Crucially, in this Maxwell-like arrangement, the strain produced by the plastic slider does not contribute to the stress production; instead, the mechanical work associated with $\gamma_p$ is entirely dissipated as heat, representing the irreversible structural reorganization of the material during flow.

Replacing Eqs. \ref{gdotds} in Eq. \ref{maxwell} we obtain the following expression for the plastic strain rate:

\begin{equation}\label{gdotdp}
\dot{\gamma}_p(t) = \frac{\sigma(t)\Fl(t)}{G_c}
\end{equation}

The order-like parameter $\theta=\Fl(t)/\dot{\gamma}(t)$ and the coupling term $\sigma\theta$ appearing in the Lyapunov function defined in the main text (Eq. 6) can be also expressed in terms of plastic strain production $d\gamma_p/d\gamma$ as follows

\begin{equation}\label{theta}
\theta = \frac{1}{\gamma_e}\frac{d\gamma_{pl}}{d\gamma},
\end{equation}

which represents the plastic strain production per unit of accumulated elastic strain, and 
\begin{equation}\label{sigmatheta}
\sigma\theta = G_c\frac{d\gamma_p}{d\gamma}.
\end{equation}

Equation~\ref{sigmatheta} is of transition-defining importance, as it reveals that the coupling term $\sigma\theta$ within the the Lyapunov function $\mathcal{V}$ is directly proportional to the system's plastic strain production, which drives the destabilization of the quiescent state, ultimately governing the emergence of a steady state characterized by a non-vanishing fluidity.

\newpage

\section{Tracking solution for $\Fl(t)$ and steady stress}
We seek a solution for constant stress $\sigma=\sigma_p$ and time dependent fluidity $\Fl(t)$ representing the fully fluidized state in oscillatory LAOS experiments. The imposed shear rate is $\dot{\gamma}=\omega\gamma_0\cos(\omega t)$. By imposing stationary stress we obtain from equation \ref{maxwell}
\begin{equation}\label{stat-stress}
\sigma_p=\frac{G_c\dot{\gamma}}{\Gamma_0+\Fl(t)}.
\end{equation}

By replacing this expression for the stress in the non linear ODE for $\Fl(t)$ (eq. 7 of the main text), we obtain the following non linear ODE for $\Fl(l)$:

\begin{equation}\label{ODEfluidity}
\dot{\Fl}=-r\Fl\left[1-\left|\frac{G_c\dot{\gamma}}{\sigma_c(\Gamma_0+\Fl(y))}\right|^{\lambda}\left|\frac{\dot{\gamma}}{\Fl}\right|^{2-\lambda}\right]
\end{equation}

Since in the fluidized state case $\Fl(t)\gg\Gamma_0$ the equation can be recast as 

\begin{equation}\label{ODEfluidity}
\dot{\Fl}=-r\Fl\left\{1-K_{2}\left[\frac{\cos{(\omega t)}}{\Fl}\right]^2\right\}
\end{equation}

with $K_{2}=\left(\omega\gamma_0\right)^{2}\left(G/\sigma_c\right)^{\lambda}$.

Equation \ref{ODEfluidity} has an analytical solution that reads:

\begin{equation}\label{fluidsol1}
\Fl(t) = \pm \sqrt{\frac{K_2}{2} \left( 1 + \frac{r^2 \cos(2\omega t) + r\omega \sin(2\omega t)}{r^2 + \omega^2} \right) + C e^{-2rt}}
\end{equation}
The constant $C$ is determined by the initial conditions, e.g $\Fl(0)=\Fl_0>0$. Since $\Fl(t)$ is a positive relaxation rate we discard the negative root. The steady solution is therefore 

\begin{equation}\label{fluidsol2}
\Fl(t) = \sqrt{\frac{K_2}{2} \left( 1 + \frac{r^2 \cos(2\omega t) + r\omega \sin(2\omega t)}{r^2 + \omega^2} \right)},
\end{equation}
that for highly dumped systems ($r\gg \omega$), reduces to 

\begin{equation}\label{fluidsol3}
\Fl(t) \simeq \sqrt{\frac{K_{2}}{r^2+\omega^2}}r|\cos(\omega t)|=\sqrt{\frac{1}{1+\frac{\omega^2}{r^2}}}\left(\frac{G_c}{\sigma_c}\right)^{\lambda}|\dot{\gamma}|
\end{equation}

Inserting \ref{fluidsol3} in \ref{stat-stress} and neglecting the small correction due to the finite $\omega/r$ ratio, we obtain an expression which links the steady stress, the shear modulus $G_c$ and the parameters arising from the BFF model, namely the characteristic stress $\sigma_c$ and the plasticity production exponent $\lambda$:

\begin{equation}\label{statstress_final}
\sigma_p=\sigma_c^{\lambda/2}G_c^{1-\lambda/2}
\end{equation}

Equations~\eqref{fluidsol3} and~\eqref{statstress_final} represent the stationary solutions for the fluidity and stress, respectively. In the context of time-dependent oscillatory flow, the term ``stationary'' refers to a \textit{tracking solution} (also known as a \textit{slow manifold} solution). According to Tikhonov's theorem \cite{tichonovDifferentialEquations1985} on singular perturbations, since the fluidity $\Fl$ acts as a fast dynamical variable, it rapidly relaxes toward a quasi-equilibrium state dictated by the instantaneous values of the slow variables, namely the stress and the strain rate. 

Specifically, these solutions represent the attractive limit cycle of the system where the fast dynamics have localized onto a lower-dimensional manifold. It can be formally verified that Eqs.~\eqref{fluidsol3} and~\eqref{statstress_final} satisfy the constraint $\psi(\sigma,\gamma,\dot{\sigma},\dot{\gamma})=1$ discussed in the main text. This constraint defines the tracking manifold of the general equation of motion for the fluidity, ensuring that the system's evolution is governed by the intrinsic bistability and symmetry-breaking mechanism rather than by transient initial conditions. In this regime, the fluidity effectively ''tracks'' the absolute value of the shear rate, leading to the observed universal scaling laws.

\newpage

\section{Extended DSS spectra from the BFF model and physical significance of $\lambda<1$}

\begin{figure}[htbp]
    \centering
    \includegraphics[width=0.8\linewidth]{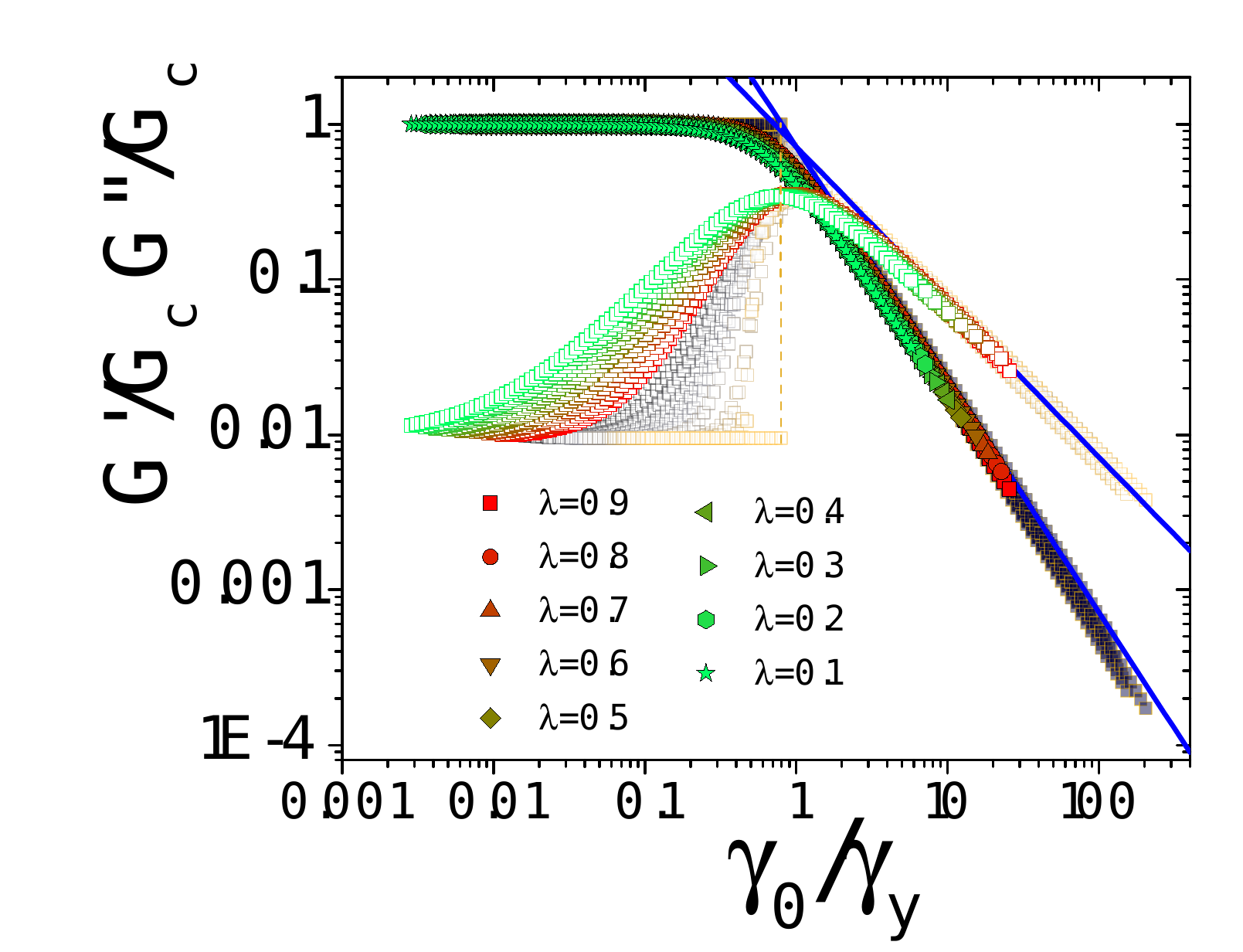}
    \caption{Scaled first-harmonic moduli simulated according to the coupled BFF equations 3-7 of the main text, for $\lambda$ varying between 0.1 (green) and 0.9 (red) and same model parameters as in Fig.~2b of the main text. The moduli are overlaid with the those reported in Fig.~2b of the main text for $1 \leq \lambda \leq 2$. Solid blue lines are the scaling laws predicted by the ERSS approximation}
    \label{fig:extendedDSS_BFF}
\end{figure}

To fully assess the boundaries of the BFF model, we we report here the DSS spectra extended 
to the parameter space where $0 \le \lambda < 1$. In this regime, the underlying 
Lyapunov potential $V(\theta)$ is strictly $\mathcal{C}^0$-continuous but exhibits 
a non-differentiable cusp at $\theta=0$. The corresponding extended scaled large-amplitude 
oscillatory shear (LAOS) spectra are presented in Fig.~\ref{fig:extendedDSS_BFF}, overlaid with the 
datasets discussed in the main text. Most importantly, setting aside potential differentiability concerns, in this interval of exponent $\lambda$ both the restoring constant $\alpha$ appearing in the 
Lyapunov function and the local curvature $\kappa$ around its minimum grow upon decreasing the linear shear modulus $G_c$. This two quantities can be computed and read:

\begin{align}
\alpha=g(\lambda)T\Delta S_{sw}^{(el)}G_c^{\lambda-1}\\
\kappa=\left.\frac{d^2\mathcal{V}}{d\theta^2}\right|_{\theta_{min}}=T\Delta S_{sw}^{(el)}G_c^{\lambda-1}\lambda(3-\lambda)
\end{align}
where $g(\lambda)=\frac{\lambda(3-\lambda)}{2-\lambda}$.\\
This paradoxical finding makes the physical significance of the regime $\lambda < 1$ currently 
questionable; its thermodynamic and mechanical validity will remain elusive until DSS spectra and intra-cycle stress responses strictly compatible with such low $\lambda$-values will be experimentally established.

\newpage

\section{Entropy production rate}
We compute the irreversible entropy production rate $\dot{S}(t)$, which quantifies the fraction of the injected energy density that is irreversibly dissipated and converted into heat \cite{hauptContinuumMechanicsTheory2002,glansdorffThermodynamicTheoryStructure1971}, as the part of the mechanical work that is not stored as recoverable elastic energy:
\begin{equation}\label{sdot1}
T\dot{S}=\sigma\dot{\gamma}-\dot{\mathcal{E}} .
\end{equation}
where $\mathcal{E}=\frac{\sigma^2}{2G_c}$ is the elastic (recoverable) energy density.
\eqref{sdot1} can be thus re-written as 
\begin{equation}\label{sdot2}
T\Sdot=\sigma\dot{\gamma}-\frac{\sigma\dot{\sigma}}{G_c},
\end{equation}
Since
\begin{equation}\label{sdot3}
\dot{\sigma}=G_c\dot{\gamma}-\sigma(\Fl+\Gamma_0),
\end{equation}
from \ref{sdot2} we get
\begin{equation}\label{sdot4}
T\Sdot=\frac{\sigma^2}{G_c}(\Fl+\Gamma_0),
\end{equation}
Hence, we can compute the entropy production expected in the linear viscoelastic (LVE) regime ($\Fl\ll\Gamma_0$) and in the fully fluidized state ($\Fl\gg\Gamma_0$) of a viscoelastic solid ($G'_l\simeq G_c\gg G''_l$).
In the LVE regime we can write from \ref{sdot4}: 
\begin{equation}\label{sdot5}
\Sdot\simeq\frac{\sigma^2}{TG_c}\Gamma_0\simeq\frac{G_c^2\gamma^2}{TG}\Gamma_0\simeq\frac{G_c\Gamma_0}{T}\gamma^2(t)
\end{equation}
In the fluidized state we get
\begin{equation}\label{sdot6}
\Sdot\simeq\frac{\sigma^2}{TG_c}\Fl\simeq\frac{\sigma_p^2}{TG_c}\Fl\simeq \frac{\sigma_p^2}{TG_c}\left(\frac{G_c}{\sigma_c}\right)^{\lambda/2}|\dot{\gamma}|=\frac{\sigma_p}{T}\omega\gamma_0\sqrt{1-\left(\frac{\gamma}{\gamma_0}\right)^2}
\end{equation}
Equations \ref{sdot5} and \ref{sdot6} have been verified numerically (see e.g fig. 3 of the main text) and describe the behavior of entropy production in the two limiting cases, far from the yielding transition.

\newpage

\section*{Maxima of the entropy production rate}

As discussed in the main manuscript, the BFF model predicts the occurrence of two distinct maxima in the entropy production rate, $\dot{S}$, within the fluidized regime. The first maximum, which is intuitively expected, arises from the peak advection occurring at the maximum shear rate. This contribution can be calculated directly by determining the apex of the ellipse described by Equation~\eqref{sdot6} in the $(\Sdot, \gamma)$ phase space. Following this approach, we obtain the first maximum value, $\dot{S}_{\text{max}}^{(1)}$, which reads:
\begin{equation}
    \Sdot_{max}^{(1)}\simeq\frac{\sigma_p^2\Fl_{max}}{TG_c}=\frac{\omega\sigma_p}{T}\gamma_0
    \label{sdot7}
\end{equation}
This analytical prediction is in excellent agreement with both the expected scaling laws and the numerical values retrieved by solving the coupled system of ordinary differential equations (ODEs) for $\sigma$ and $\Fl$ in the highly damped limit ($r \gg \omega$), as illustrated by the representative example in Fig. 3 of the main text.

Less trivial is the analytical derivation of a closed-form expression for the secondary maximum of the entropy production rate, which stems from the stress inversion observed at the flow reversal points. Numerical simulations reveal that this secondary peak grows with a power-law exponent strictly lower than unity, typically remaining very close or identical to $1/2$ (see Fig.~3 of the main text). To gain analytical insight, we can approximate the fluidity evolution in the vicinity of the flow reversal points by adopting the ERSS approximation (Eqs.~\ref{approxgammadot}-\ref{approxstress}). Within this framework, the fluidity can be explicitly expressed as:
\begin{equation}
    \Fl(\tau) = \gamma_0\omega^2\tau\frac{1}{\sigma_c^{\frac{\lambda}{2-\lambda}}}\left|\frac{1}{2}G_c\omega^2\gamma_0\tau^2-\sigma_p\right|^{\frac{\lambda}{2-\lambda}}
    \label{eq:fluidity_erss}
\end{equation}

By leveraging this formulation, we can calculate the characteristic time $\tau_m$ (measured from the flow reversal instance where $\dot{\gamma}=0$) at which the local maximum of the fluidity occurs. By imposing $d\Fl(\tau)/d\tau=0$ we find:
\begin{equation}
    \tau_m = \sqrt{\frac{2\sigma_p(2-\lambda)}{G_c\omega^2\gamma_0(2+\lambda)}}
    \label{eq:tau_m}
\end{equation}
Consequently, an approximate analytical expression for the secondary maximum of $\dot{S}$ can be obtained by evaluating both the stress and the fluidity at $\tau = \tau_m$ and substituting them back into Equation~\eqref{sdot4}. This procedure yields the following closed-form result:
\begin{multline}
    \dot{S}_{\text{max}}^{(2)} \simeq \frac{\sigma^2(\tau_{\max})}{TG_c}\left[\Fl(\tau_{max})+\Gamma_0\right)]=\\\frac{1}{TG_c}\left(\frac{2\lambda}{2+\lambda}\sigma_p\right)^2\left[\frac{\gamma_0\omega^2}{\sigma_c^{\frac{\lambda}{2-\lambda}}}\left(\frac{2\lambda}{2+\lambda}\sigma_p\right)^{\frac{\lambda}{2-\lambda}}\sqrt{\frac{2\sigma_p(2-\lambda)}{G_c\omega^2\gamma_0(2+\lambda)}}+\Gamma_0\right]
    \label{eq:Smax2}
\end{multline}
The ERSS approximation thus predicts a secondary maximum that scales as $\gamma_0^{1/2}$, capturing very-well the power-law behavior observed numerically in the full BFF model. At this stage, it is particularly interesting to observe that this secondary peak obeys a scaling law, $\dot{S}_{\text{max}}^{(2)} \sim \gamma_0^{1/2}$, which is inversely proportional to the characteristic stress reversal time, $\tau_d \sim \gamma_0^{-1/2}$. As will be detailed in Sections~\ref{Sw_entropy}-\ref{switchelastic}, this reciprocal scaling has profound implications for the total entropy produced during the stress reversal process. Specifically, when the ERSS approximation and the large-amplitude limit are taken into account, it leads to a constant, amplitude-independent value of the switching entropy, which is the prefactor ruling the magnitude of the total entropy produced by the system during the transition from one steady-state flow ($+\dot{\gamma}$) configuration to the opposing one ($-\dot{\gamma}$) (Eq. \ref{switching_final}).





\newpage

\section{Switching entropy}\label{Sw_entropy}
From Equation~\eqref{sdot4}, the entropy production per unit volume during the switching process—defined by the transition between a generic stress state $-\sigma \to \sigma$— is given by:
\begin{equation}\label{switching1}
\Delta S_{sw}(\sigma) = \frac{1}{T G_c} \int_{\tau_{-\sigma}}^{\tau_{\sigma}} \sigma^2(\tau) \left[ \Fl(\tau) + \Gamma_0 \right] d\tau
\end{equation}
where $\tau_{-\sigma}$ and $\tau_{\sigma}$ denote the time instances at which the stress assumes the values $-\sigma$ and $\sigma$, respectively. 

As previously discussed, for highly damped systems ($r \to \infty$), the fluidity evolves along the manifold $\psi(\sigma, \gamma, \dot{\sigma}, \dot{\gamma}) = 1$, which yields:
\begin{equation}\label{fluid1}
\Fl = |\dot{\gamma}| \left| \frac{\sigma}{\sigma_c} \right|^{\frac{\lambda}{2-\lambda}}
\end{equation}

To evaluate the integral in Equation~\eqref{switching1}, it is convenient to change the integration variable from time to stress:
\begin{equation}\label{change_var}
d\tau = \frac{d\sigma}{\dot{\sigma}} = \frac{d\sigma}{G_c \dot{\gamma} - \sigma(t) \left( \left| \frac{\sigma(t)}{\sigma_c} \right|^n + \Gamma_0 \right)}
\end{equation}
where $n = \lambda/(2-\lambda)$. Here, we consider a stress reversal from negative to positive values under a strictly positive shear rate, $\dot{\gamma} > 0$. The change of variables from time to stress introduces the dynamical measure $d\sigma/\dot{\sigma}$, which weights each stress value by the time spent along the trajectory.
Substituting Eqs~\eqref{change_var} and \eqref{fluid1} into eq. \eqref{switching1} leads to the following expression for the switching entropy:
\begin{multline}\label{switching2a}
T \Delta S_{sw}(\sigma) = T\int_{-\sigma}^{+\sigma} \dot{S}(\tilde{\sigma}) \frac{d\tilde{\sigma}}{\dot{\tilde{\sigma}}} = \frac{1}{G_c} \int_{-\sigma}^{+\sigma} \frac{|\tilde{\sigma}|^{n+2} + \frac{\tilde{\sigma}^2 \Gamma_0}{\dot{\gamma}} |\sigma_c|^n}{G_c |\sigma_c|^n - \tilde{\sigma} |\tilde{\sigma}|^n - \frac{\tilde{\sigma} \Gamma_0}{\dot{\gamma}} |\sigma_c|^n} d\tilde{\sigma} \\
= \frac{1}{G_c} \int_{-\sigma}^{+\sigma} \frac{|\tilde{\sigma}|^{n+2} + \frac{\tilde{\sigma}^2 \Gamma_0}{\dot{\gamma}} |\sigma_c|^n}{\sigma_p^{n+1} - \tilde{\sigma} |\tilde{\sigma}|^n - \frac{\tilde{\sigma} \Gamma_0}{\dot{\gamma}} |\sigma_c|^n} d\tilde{\sigma}
\end{multline}
where we have invoked Equation~\eqref{statstress_final} to utilize the identity $G_c |\sigma_c|^n = \sigma_p^{n+1}$. 

In the large-amplitude limit, $\frac{\Gamma_0}{\gamma_0 \omega} \to 0$, the contribution from spontaneous relaxation becomes negligible near $\sigma = 0$, where $\dot{\gamma} \neq 0$. Consequently, we focus on the switching entropy purely associated with plastic strain production. By accounting for the change in the sign of the stress, the integrals in Equation~\eqref{switching2a} can be recast as:
\begin{equation}\label{switching2b}
T \Delta S_{sw}(\sigma) = \frac{1}{G_c} \int_0^{\sigma} \left( \frac{\tilde{\sigma}^{n+2}}{\sigma_p^{n+1} + \tilde{\sigma}^{n+1}} + \frac{\tilde{\sigma}^{n+2}}{\sigma_p^{n+1} - \tilde{\sigma}^{n+1}} \right) d\tilde{\sigma}
\end{equation}

The integral in equation \eqref{switching2b} can be computed analytically and gives equation 12 of the main text that we report here for the sake of clarity
\begin{equation}\label{switching_final}
T\Delta S_{sw}(z)=f(\lambda)\,\mathcal{E}_{el}\,z^{\frac{4}{f(\lambda)}}\,{}_{2}F_{1}\left[1,\frac{1}{2-f(\lambda)},\frac{3-f(\lambda)}{2-f(\lambda)},z^{4\left(\frac{2}{f(\lambda)}-1\right)}\right]
\end{equation}
where $f(\lambda)=\frac{2(2-\lambda)}{3-\lambda}$, $z=\frac{\sigma}{\sigma_p}$, ${}_{2}F_{1}(a,b,c,z)$ the Gaussian hypergeometric function, and $\mathcal{E}_{el}=\sigma_p^2/(2G_c)$ is the energy stored under steady stress. 
Therefore, the dissipation power associated with the stress reversal from $-\sigma$ to $+\sigma$ scales, for small values of $z$, as a power law of the form $T\Delta S_{sw} \sim f(\lambda) \mathcal{E}_{el} z^{\frac{4}{f(\lambda)}}$. 
Consequently, a system characterized by a higher viscoplastic fragility (i.e., large $\lambda$) is associated—at a constant recoverable elastic energy in the steady stress state—with a steeper and more localized entropy production within the stress domain, leading up to the singular behavior reached as $\lambda \to 2$, as shown in figure 3g of the main manuscript. 
Finally, the dissipated power $T\Delta S_{sw}$ diverges logarithmically as $z \to 1$, owing to the vanishing behavior of the residence-time measure in the stress domain, defined by $d\sigma/d\dot{\sigma}$.

\newpage

\section{Elastic switching entropy}\label{switchelastic}
Remarkably the prefactor (or amplitude) in equation \eqref{switching_final} is the value of the switching entropy due only to the elastic recoil, namely the entropy that can be derived without accounting for stress relaxation during switching. This obtained by neglecting the terms $\tilde{\sigma}^{n+1}$ at the denominators of the integrand function. In this case equation \eqref{switching2b} becomes 

\begin{multline}\label{switching3}
T \Delta S_{sw}^{(el)}(\sigma) = \frac{2}{G_c \sigma_p^{n+1}} \int_0^{\sigma} \tilde{\sigma}^{n+2} d\tilde{\sigma}=\frac{2}{(n+3)G_c\sigma_p^{n+1}}\sigma^{n+3}=\\\frac{4}{n+3}\frac{\sigma_p^2}{2G_c}z^{n+3}=f(\lambda)\mathcal{E}_{el}z^{4/f(\lambda)}
\end{multline}
which reach the value 
\begin{equation}\label{switching3}
T \Delta S_{sw}^{(el)}(\sigma_p) =f(\lambda)\mathcal{E}_{el}
\end{equation}
for the full switching between the two steady states, and represents therefore a material characteristic constant.\\

The very same result can be obtained in the time domain if we consider the elastic ERSS approximation (Eqs. \ref{approxgammadot}-\ref{approxstress}), as we detail hereafter.

In the dumped limit $r\rightarrow\infty$ and $\dot{\gamma}>0$ equation \eqref{fluid1} for the fluidity becomes 
\begin{equation}\label{ediffFlambda}
\Fl(\tau)=\gamma_0\omega^2\tau\frac{1}{\sigma_c^{\frac{\lambda}{2-\lambda}}}\left|\frac{1}{2}G_c\omega^2\gamma_0 \tau^2-\sigma_p\right|^{\frac{\lambda}{2-\lambda}}
\end{equation}
Hence, the elastic switching entropy can be computed as the large amplitude limit of the integrated entropy production rate
\begin{multline}\label{switchentropylambda}
T\Delta S^{(el)}_{sw}=\lim_{\gamma_0\rightarrow\infty}\int_0^{\tau_d}T\Sdot(\tau)d\tau=\\\lim_{\gamma_0\rightarrow\infty}\frac{1}{G_c}\int_0^{\tau_d}\left(\frac{1}{2}G_c\omega^2\gamma_0\tau^2-\sigma_p\right)^2\gamma_0\omega^2\tau\frac{1}{\sigma_c^{\frac{\lambda}{2-\lambda}}}\left|\frac{1}{2}G_c\omega^2\gamma_0 \tau^2-\sigma_p\right|^{\frac{\lambda}{2-\lambda}}d\tau
\end{multline}
where $\tau_d$ is the switching time and is given by equation \eqref{switchingtime}.
We find
\begin{equation}\label{switching_generic}
T\Delta S^{(el)}_{sw}=\frac{1}{2}f(\lambda)\sigma_c^{\lambda}G_c^{1-\lambda}
\end{equation}
with $f(\lambda)=\frac{2(2-\lambda)}{3-\lambda}$, which coincides with Eq. \eqref{switching3}, bearing in mind that $\sigma_p=\sigma_c^{\lambda/2}G_c^{1-\lambda/2}$.

\newpage

\section{Decay exponents of the first-harmonic moduli (Data shown in fig. 1 - a,b)}
\begin{figure}[htbp]
    \centering
    \includegraphics[width=0.75\linewidth]{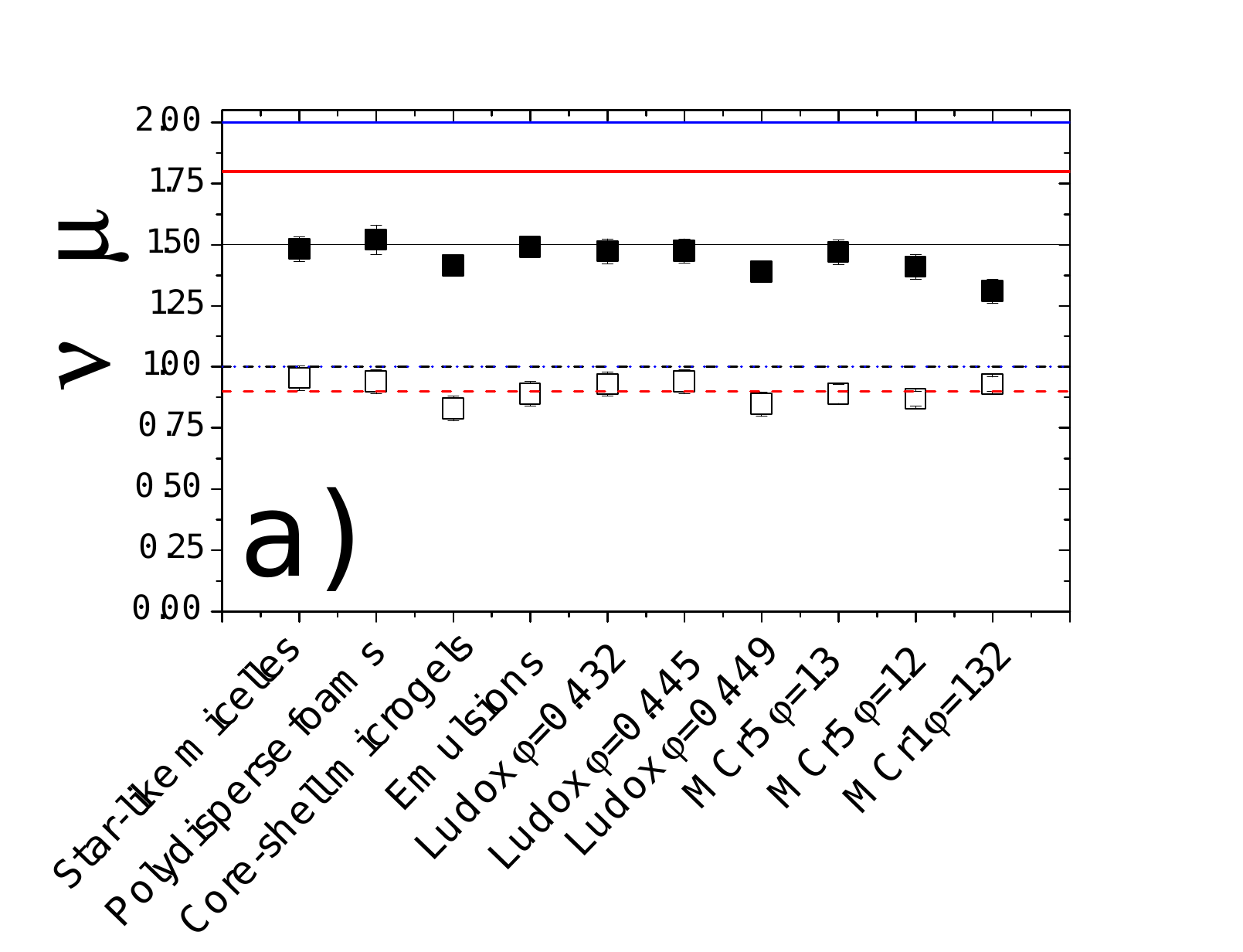}
    \includegraphics[width=0.75\linewidth]{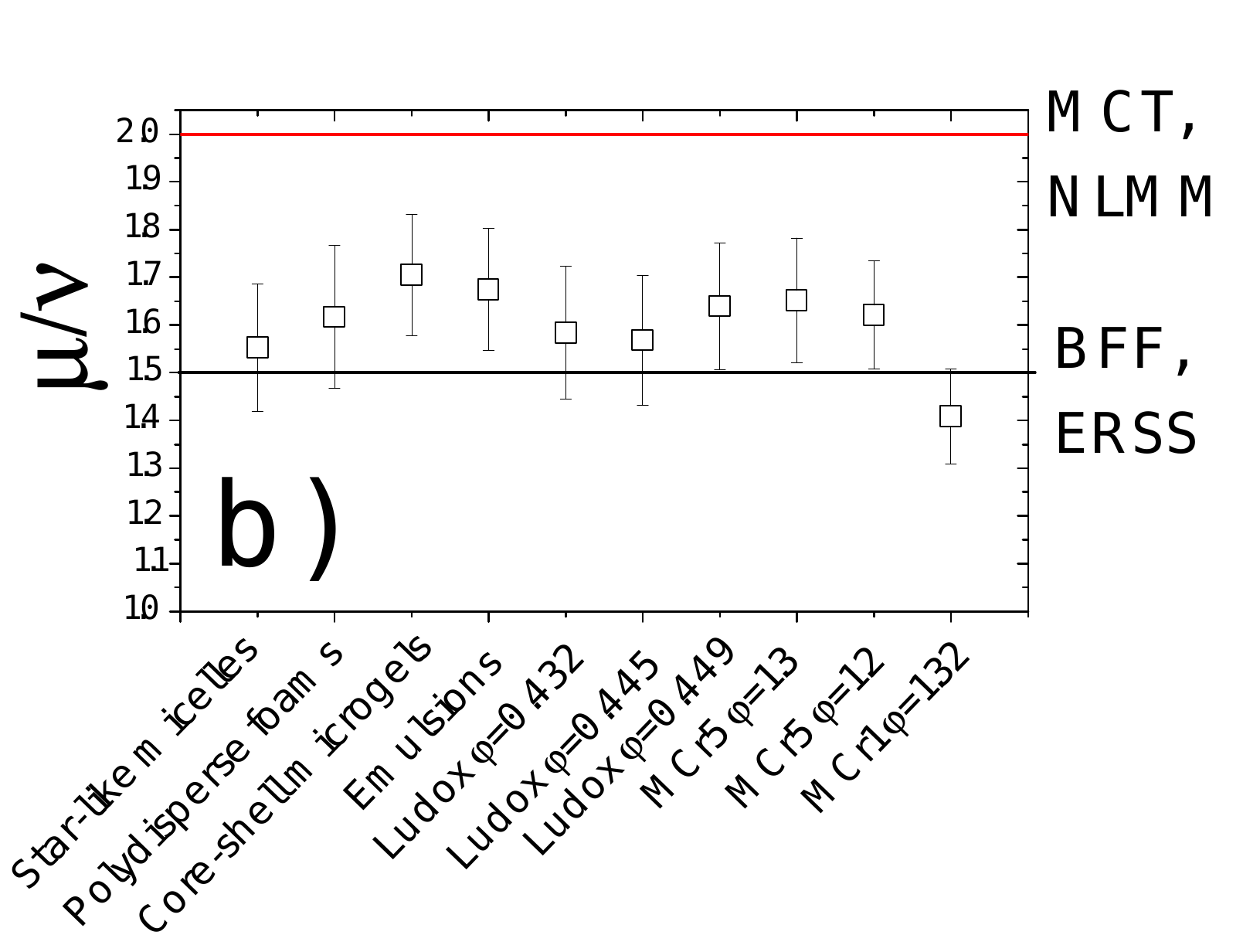}
    \caption{Power-law exponents a) and their ratio b) of the first-harmonic moduli obtained in the LAOS regime for the systems shown in figure 1 of the main text. In panel a) dashed lines and solid lines mark the exponents predicted by MCT (red)\cite{miyazakiNonlinearViscoelasticityMetastable2006}, non-linear Maxwell model (NLMM) (blue) \cite{koumakisDirectComparisonRheology2012} and the BFF/ERSS model (black) discussed in the present work. In panel b) solid lines show the exponent ratio $\mu/\nu$ predicted by MCT and NLMM (red) and BFF/ERSS model (black) }
    \label{fig:exponents}
\end{figure}

We report here the exponents with errors obtained for the moduli decay of the systems shown in fig. 1 of the main text. The exponents have been obtained by fitting the last half decade of amplitudes probed by dynamic strain sweeps. In Figure \ref{fig:exponents}-a we report these exponents compared to the theoretical scaling predicted by the ERSS approximation and, by solving numerically the the BFF model for any plasticity production exponent $\lambda$ (fig. 2 of the main text): $G'\sim\gamma_0^{-\mu}$ and $G''\sim\gamma_0^{-\nu}$, with $\mu=1.5$ and $\nu=1.0$\footnote{The mean values and standard deviations of the exponents extracted from power-law fits to the first-harmonic moduli computed via the BFF model for $1.0<\lambda<1.975$ are $\mu=1.51\pm 0.02$ and $\nu=0.97\pm 0.02$}. We further compare them with previous predictions from MCT ($\mu=1.8$, $\nu=0.9$) and non-linear Maxwell scaling ($\mu=2$, $\nu=1$)\cite{miyazakiNonlinearViscoelasticityMetastable2006}. In Figure \ref{fig:exponents}-b reports the ratio $\mu/\nu$ for the same systems shown in panel a) compared to the one predicted by ERSS and BFF model $\mu/\nu=1.5$ (present work) and the ratio $\mu/\nu=2.0$. Finally, we recall that the Soft Glassy Rheology (SGR) model \cite{sollichRheologicalConstitutiveEquation1998} does not predict universal scalings of the moduli. 

\newpage

\section{Elastoviscoplastic parameters explored in figure 2-c}
\begin{figure}[htbp]
    \centering
    \includegraphics[width=1\linewidth]{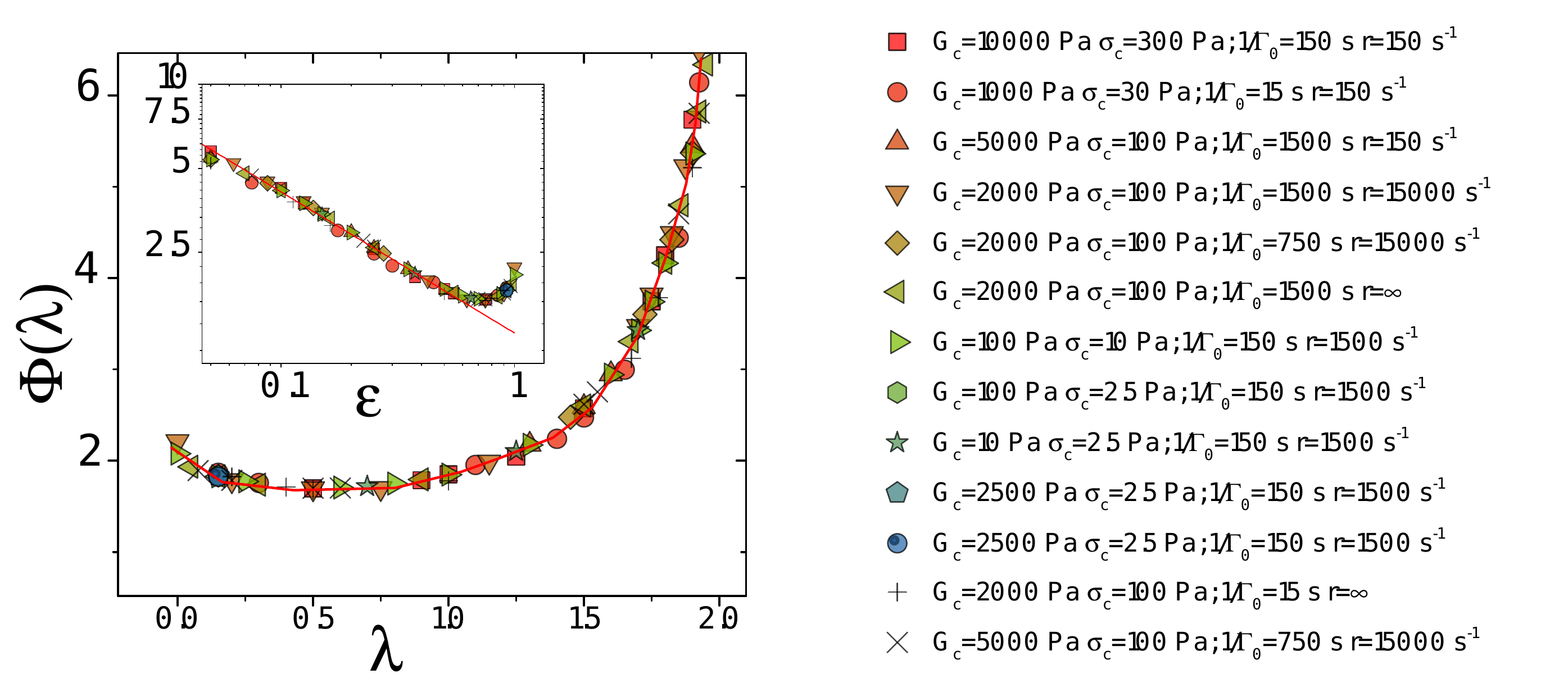}
    \caption{
    \textbf{Robustness of the viscoplastic fragility $\Phi(\lambda)$ across diverse elastoviscoplastic parameters.} 
    This figure supplements the numerical results shown in Fig.~2c of the main manuscript, showing the evolution of the viscoplastic fragility curve, $\Phi(\lambda)$, over a wide range of constitutive parameters, shown here in the legend. The explored parameter space spans more-than-one order of magnitude for each parameter:
        \protect\\ $\bullet$ Elastic modulus: $G_c \in [10, 5000]$~Pa
        \protect\\ $\bullet$ Critical stress: $\sigma_c \in [2.5, 300]$~Pa
        \protect\\ $\bullet$ Characteristic spontaneous relaxation time: $1/\Gamma_0 \in [15, 1500]$~s
        \protect\\ $\bullet$ Dumping coefficient: $r \in [150, 15000]$~s$^{-1}$ (including the infinite dumping limit $r = \infty$)
    }
    \label{fig:fragility}
\end{figure}

\newpage

\section{Viscoplastic fragility from dynamic strain sweeps}
We present below the result of analysis of the loss modulus curves, $G''(\gamma_0)$, from which the viscoplastic fragility parameter, $\Phi$, was extracted. Figure~\ref{fig:fragility}-a,b,c illustrates a representative example for three distinct curves where $\Phi$ was computed directly from the viscoelastic moduli. Specifically, the experimental data were fitted from the linear regime up to the local maximum using an auxiliary Boltzmann-type function of the form:
\begin{equation}
    \frac{\Delta G''(\gamma_0)}{\Delta G''_M} = A_2 + \frac{A_2-A_1}{1+\exp\left[\frac{1}{\gamma_M}(\gamma_0-\gamma_0^*)/dx\right]},
    \label{eq:boltzmann_fit}
\end{equation}
where $A_2$ and $A_1$ are the values of the two plateaus of the sigmoid, $\gamma_0^*$ is the value of the strain amplitude at the inflection point, where $\frac{\Delta G''(\gamma_0)}{\Delta G''_M}$ is maximum, and $dx$ is the slope factor. 
Under this choice, the viscoplastic fragility is analytically obtained as:
\begin{equation}
    \Phi = \frac{A_2 - A_1}{4 \cdot dx}
    \label{eq:phi_definition}
\end{equation}
The $\Phi$ values obtained for the diverse systems under study were subsequently mapped onto the universal $\Phi(\lambda)$ master curve predicted by the BFF model (reported in Fig.~2 of the main text) by manually adjusting the plasticity production exponent $\lambda$ (Figure~\ref{fig:fragility}-d). Crucially, for all investigated systems, we systematically find a value of $\lambda \geq 1$. 

It is worth emphasizing that a dedicated, comprehensive study focusing on the viscoplastic fragility derived from DSS spectra, as defined in the present framework, is thoroughly in order. In particular, a higher experimental resolution of the loss modulus behavior in the immediate vicinity of the yielding transition is strictly necessary to evaluate the maximum derivative of its pre-yielding increment with enhanced precision.

\begin{figure}[htbp]
    \centering
    \includegraphics[width=1.0\linewidth]{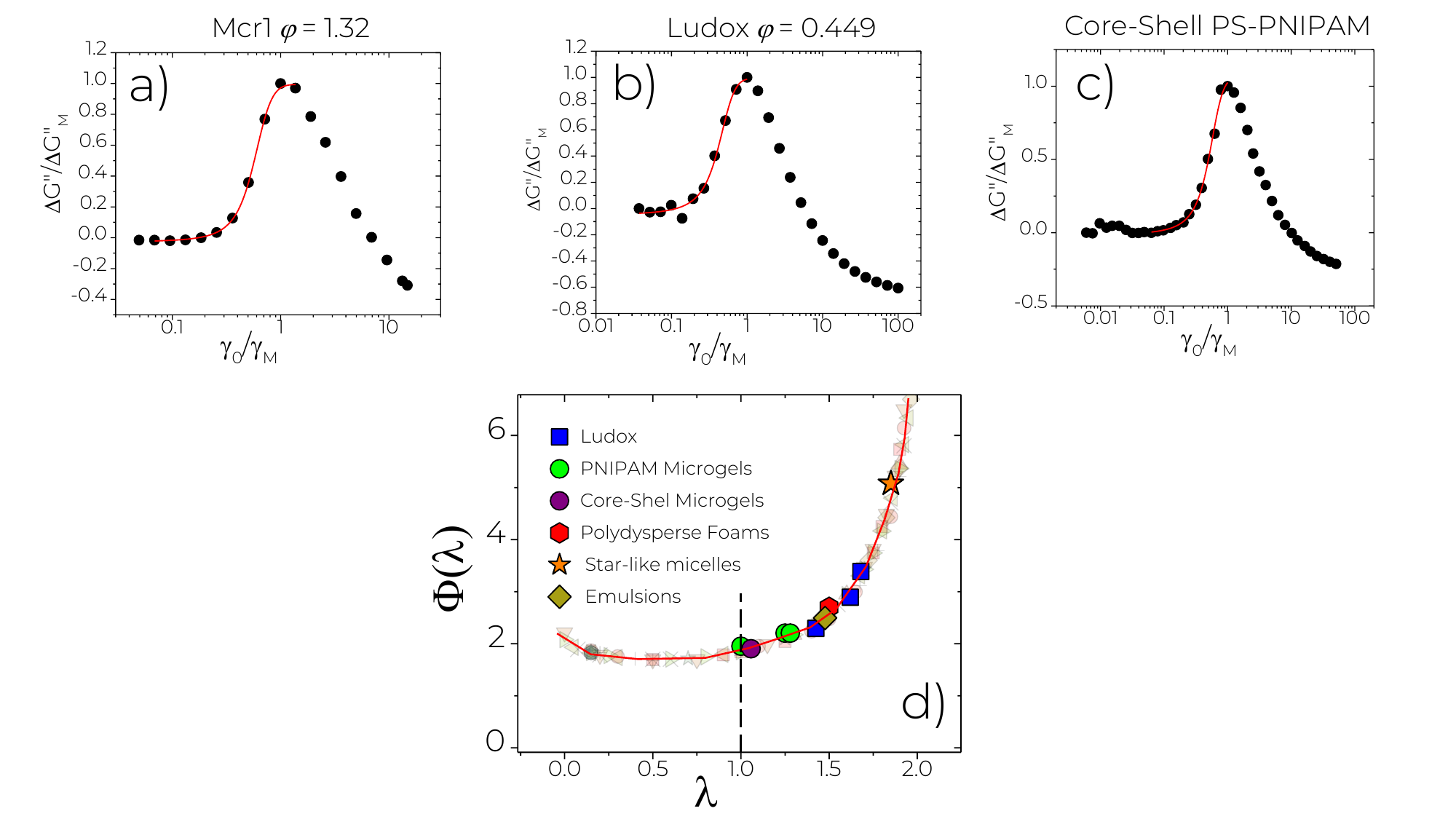}
    \caption{a) b) c): Normalized loss modulus increment $\Delta G''/\Delta G''_M$ for three representative experimental systems shown in Fig. 1 of the main text as a function of the scaled strain amplitude $\gamma_0/\gamma_M$. The red solid line is a Boltzmann fit of the data to extract the viscoplastic fragility as the maximum derivative of $\Delta G''/\Delta G''_M$. d) Viscoplastic fragility obtained for the 10 systems reported in figure 1 of the main text superimposed to the theoretical mastercurve predicted by the BFF model. The $\lambda$-values have been adjusted manually to superimpose each value to the mastercurve $\Phi(\lambda)$ predicted by the BFF model.}  
    \label{fig:fragility}
\end{figure}

\newpage

\section{Minimum viscoplastic fragility}
\begin{figure}[htbp]
    \centering
    \includegraphics[width=0.75\linewidth]{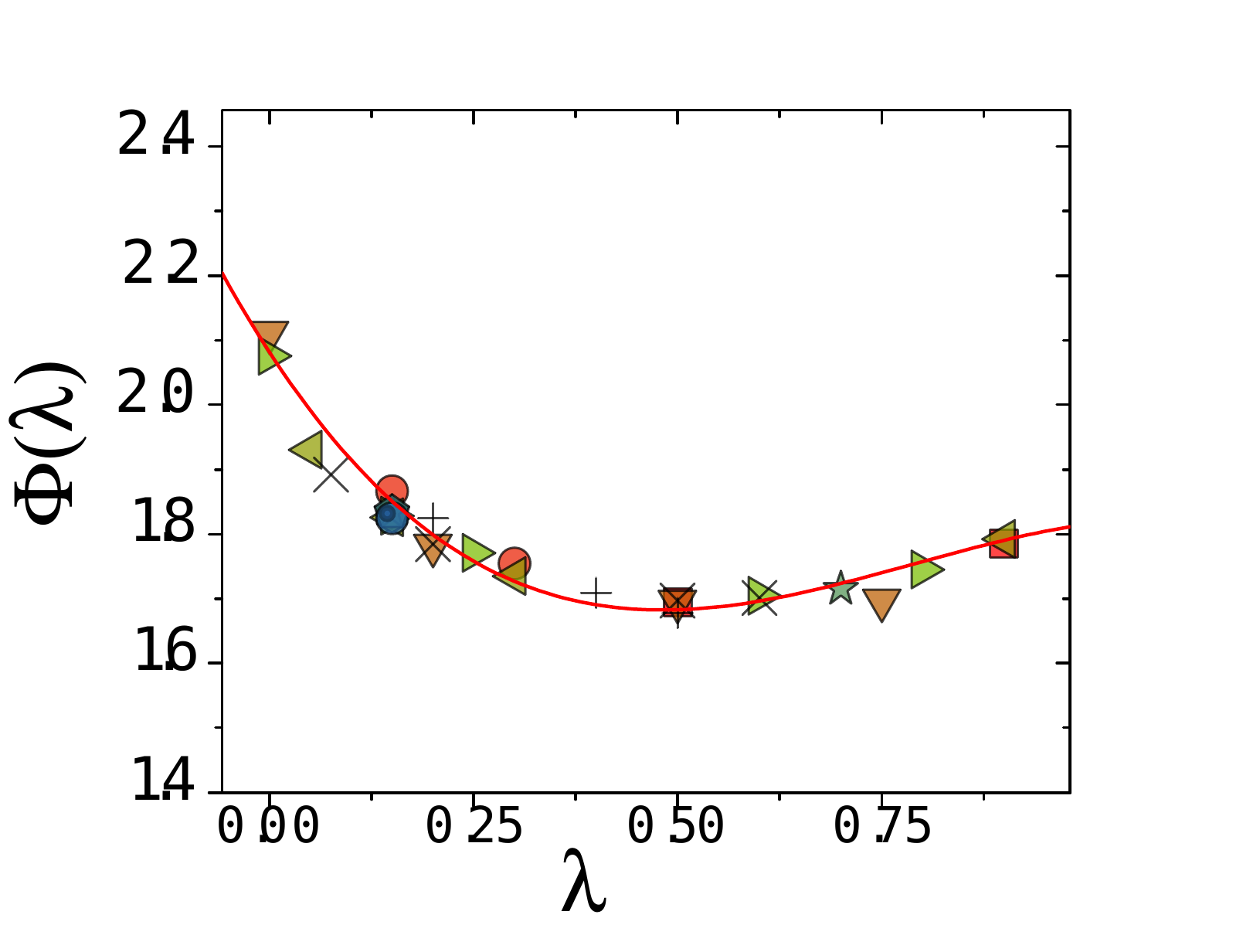}
    \caption{\textbf{Extraction of the minimum position and threshold value of the viscoplastic fragility $\Phi(\lambda)$.} 
    The localization of the critical parameters had been achieved by fitting the cubic auxiliary polynomial function $f(x) = A + Bx + Cx^2 + Dx^3$ to the computed fragilities in the rage $0\leq\lambda\leq1$. The non-linear least-squares regression yielded the following optimal values: 
    $A = 2.08 \pm 0.02$, 
    $B = -1.94 \pm 0.19$, 
    $C = 2.90 \pm 0.45$, and 
    $D = -1.22 \pm 0.29$. 
    We obtained the fragility minimum positioned at a critical exponent of $\lambda_{\text{min}} = 0.491\pm 0.024$, corresponding to a minimum viscoplastic fragility value of $\Phi(\lambda_{\text{min}}) = 1.674\pm 0.038$.}  
    \label{fig:fragility}
\end{figure}

\newpage

\section{Analytical Derivation of the First-Harmonic Moduli}

We derive the expressions for the first-harmonic viscoelastic moduli, $G'(\gamma_0)$ and $G''(\gamma_0)$, directly from the Maxwell constitutive equation \eqref{maxwell}. By projecting the stress evolution onto the orthogonal harmonic basis of the strain and strain rate, we can isolate the individual viscous, and plastic contributions to the cycle-averaged dissipation $\D(\gamma_0)=\pi\gamma_0^2G''(\gamma_0)$.

We thus begin by evaluating the loss modulus, $G''$. Multiplying the constitutive Maxwell equation \eqref{maxwell} by the instantaneous strain $\gamma(t)$ and integrating over a full oscillation period $T = 2\pi/\omega$ yields:
\begin{equation}\label{inter1}
\int_{0}^{T}\dot{\sigma}\gamma \, dt = -\int_0^{T}\left[\Gamma_0+\Fl(t)\right]\sigma\gamma \, dt
\end{equation}
where we have exploited the orthogonality relation $\int_{0}^{T}\dot{\gamma}\gamma \, dt = 0$. Integration by parts of the left-hand side, imposing the periodic boundary conditions $\sigma(T)=\sigma(0)$ and $\gamma(T)=\gamma(0)$, leads to:
\begin{equation}\label{inter2}
-\int_{0}^{T}\dot{\gamma}\sigma \, dt = -\int_0^{T}\left[\Gamma_0+\Fl(t)\right]\sigma\gamma \, dt
\end{equation}
The left-hand side of Equation~\eqref{inter2} represents the total mechanical energy dissipated per unit volume over a single cycle, $\D$. Consequently, the loss modulus can be directly linked to the time-dependent fluidity according to:
\begin{equation}
G''(\gamma_0) = \frac{1}{\pi\gamma_0}\int_0^T \sin(\omega t)\left[\Fl(t)+\Gamma_0\right]\sigma \, dt = \frac{\Gamma_0 G'}{\omega} + \frac{1}{\pi\gamma_0}\int_0^T \sin(\omega t)\sigma \Fl(t) \, dt
\end{equation}
By adopting the standard notation for the $L^2$ inner product, $\langle f, g \rangle = \frac{\omega}{2\pi} \int_0^{2\pi/\omega} f(t) \overline{g(t)} \, dt$, this expression can be compactly recast as:
\begin{equation}
G'' = \frac{2}{\gamma_0^2\omega} \left[ \langle\sigma\gamma, \Gamma_0\rangle + \langle\sigma\gamma, \Fl(t)\rangle \right] = \frac{2\Gamma_0}{\gamma_0^2\omega} \left[ \langle\sigma, \gamma\rangle + \left\langle\sigma, \gamma \frac{d\gamma_{pl}}{d\gamma_{v}}\right\rangle \right]
\end{equation}
The second term inside the brackets accounts for the non-linear plastic dissipation and exhibits a distinct local maximum as a function of the strain amplitude. Recalling that the infinitesimal plastic work dissipated per unit volume is given by $dW_{pl}=\sigma d\gamma_{pl}$, the loss modulus can be rewritten as:
\begin{equation}\label{G2}
G'' = \frac{2\Gamma_0}{\gamma_0^2\omega} \left[ \langle\sigma,\gamma\rangle + \left\langle\frac{dW_{pl}}{d\gamma_{v}},\gamma\right\rangle \right] = \frac{2\Gamma_0}{\gamma_0^2\omega} \left[ \langle\sigma,\gamma\rangle + \eta_0 \langle \dot{\gamma}_p,\gamma\rangle \right]
\end{equation}
where the viscous scaling stems from the identities $\frac{dW_{pl}}{d\gamma_{v}} = \sigma \frac{\dot{\gamma}_p}{\dot{\gamma}_v}$ and $\sigma = \eta_0 \dot{\gamma}_v$. 

Crucially, the coupling between the instantaneous stress and the fluidity field directly governs the plastic strain rate through the relation \eqref{gdotdp}:
\begin{equation}\label{plasticrate}
\dot{\gamma}_p = \frac{\sigma}{G_c}\Fl
\end{equation}
which provides a direct route to compute the irreversible deformation rate.

Equation~\eqref{G2} can be further expanded by normalizing the kinematic variables - strain and strain rate - by their respective amplitudes. Defining the dimensionless normalized strain as $\hat{\gamma}(t) = \sin(\omega t)$ and the normalized strain rate as $\hat{\dot{\gamma}}(t) = \cos(\omega t)$, we obtain:
\begin{align}\label{G2_b}
G'' (\gamma_0) &= \frac{2\Gamma_0}{\gamma_0^2\omega} \left[ \langle\sigma,\gamma\rangle + \eta_0\langle \dot{\gamma}_p,\gamma\rangle \right] = \frac{2\Gamma_0}{\gamma_0^2\omega} \left[ \gamma_0 \langle\sigma,\hat{\gamma}\rangle + \eta_0\gamma_0\langle \dot{\gamma}_p,\hat{\gamma}\rangle \right] \nonumber \\
&= \frac{2\Gamma_0}{\gamma_0\omega} \left[ \langle\sigma,\hat{\gamma}\rangle + \eta_0\langle \dot{\gamma}_p,\hat{\gamma}\rangle \right] = \frac{2 G_c}{\gamma_0\omega} \left[ \langle\dot{\gamma}_v,\hat{\gamma}\rangle + \langle \dot{\gamma}_p,\hat{\gamma}\rangle \right]=G''_v(\gamma_0)+G''_p(\gamma_0)
\end{align}
This formulation explicitly partitions the total linear and non-linear dissipation into the underlying viscous ($\dot{\gamma}_v$) and plastic ($\dot{\gamma}_p$) shear rate components.

Similarly, by multiplying the stress derivative $\dot{\sigma}$ by the strain rate $\dot{\gamma}$ and integrating by parts following the same procedure detailed in equations~\eqref{inter1} and~\eqref{inter2}, the first-harmonic storage modulus $G'$ can be expressed as:
\begin{align}\label{G1_a}
G'(\gamma_0) = G_c \left[ 1 - \left( \frac{2}{\omega\gamma_0}\langle\dot{\gamma}_v, \hat{\dot{\gamma}}\rangle + \frac{2}{\omega\gamma_0}\langle\dot{\gamma}_p, \hat{\dot{\gamma}}\rangle \right) \right]=G_c\left[1-G_v'(\gamma_0)-G_p'\right(\gamma_0)]
\end{align}

The relative viscoelastic (recoverable) and plastic (non-recoverable) dissipative contributions to the first-harmonic moduli are shown in Fig.~\ref{fig:plastic1} for a representative case with $\lambda = 1.25$. As analytically anticipated, the individual components exhibit a fundamentally distinct coupling within the linear and non-linear regimes. In the case of the loss modulus, $G''(\gamma_0)$, the viscous and plastic shear rate contributions are strictly additive, and their sum directly reconstructs the total macroscopic energy dissipation cycle by cycle. Conversely, the evolution of the storage modulus, $G'(\gamma_0)$, is progressively reduced by the subtraction of both viscous and plastic rate projection onto the total strain rate, thereby dictating the characteristic non-linear decay of the storage modulus.

Equations \eqref{G1_a} and \eqref{G2_b} have been employed to compute the first-harmonic moduli after having obtained $\sigma(t)$ and $\Fl(t)$ from the numerical integration of the coupled ODEs 3 and 7 of the main manuscript.  

\newpage

\section{Plastic loss modulus $G''_p$}

\begin{figure}[htbp]
    \centering
    \includegraphics[width=1.0\linewidth]{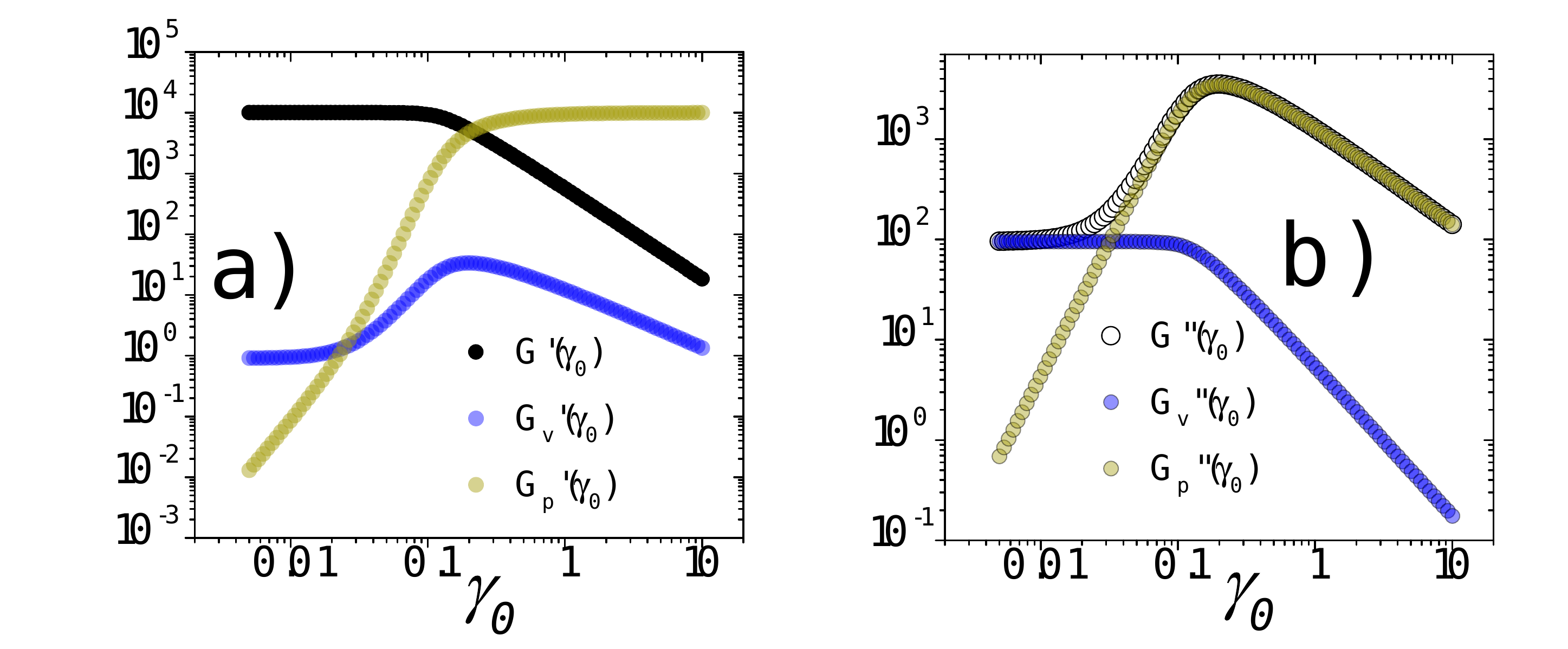}
    \caption{Viscoplastic decomposition of the first-harmonic viscoelastic moduli.
    The individual viscous and plastic dissipative contribution to both the storage modulus \textbf{(a)} and loss modulus \textbf{(b)} are displayed alongside the total macroscopic response. The corresponding constitutive components — $G'_v(\gamma_0)$, $G'_p(\gamma_0)$, $G''_v(\gamma_0)$, and $G''_p(\gamma_0)$ — have been computed for a representative model parameters: $\lambda = 1.25$, $G_c = 10000$ Pa, $r = 1500$ s$^{-1}$, $\Gamma_0 = 0.0067$ s$^{-1}$, $\sigma_c = 300$ Pa, and angular frequency of $\omega = 0.7$~rad/s.}
    \label{fig:viscoplastic_decomposition}  
    \label{fig:plastic1}
\end{figure}

\begin{figure}[htbp]
    \centering
    \includegraphics[width=1.0\linewidth]{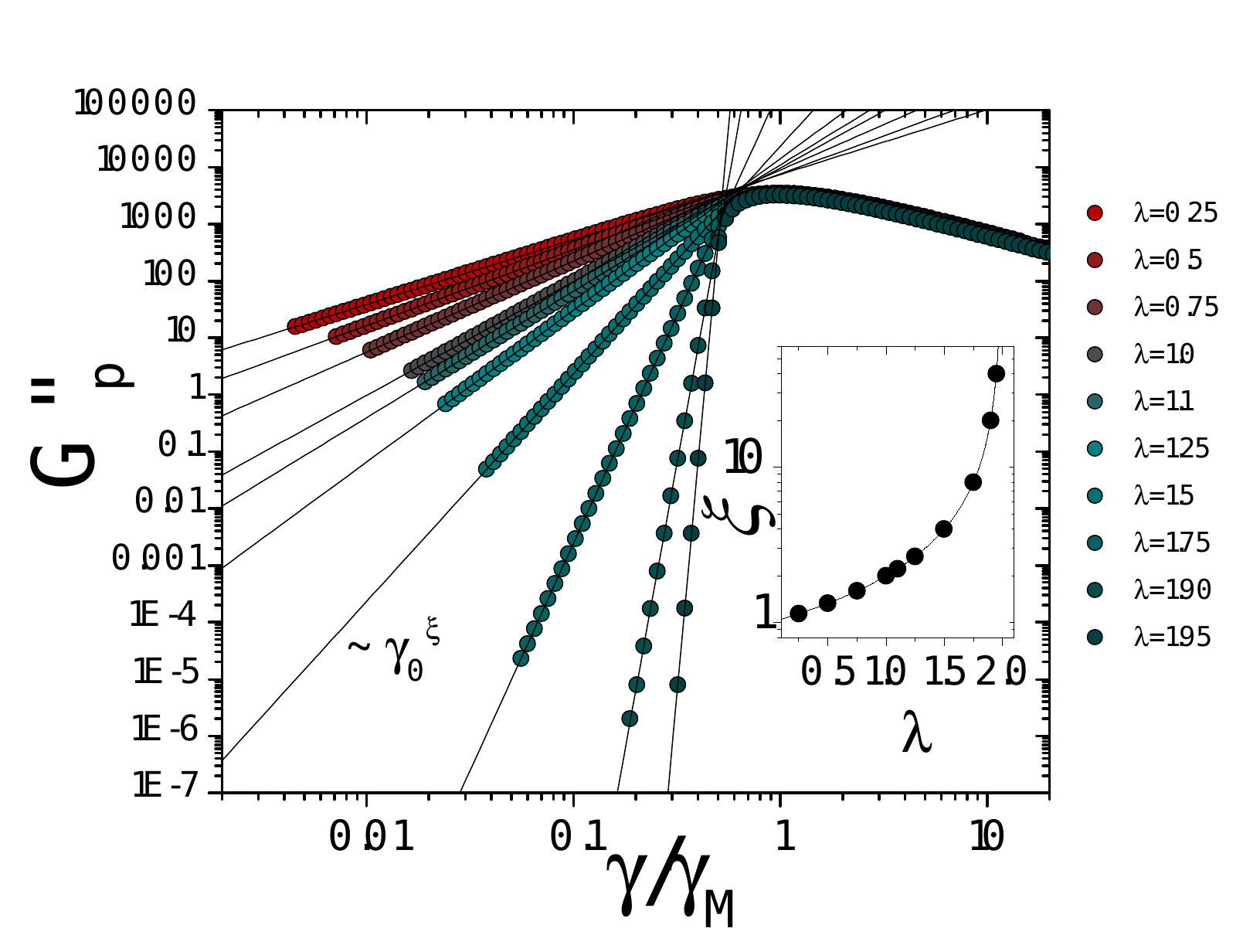}
    \caption{\textbf{Scaling behavior and exponent universality of the plastic loss modulus.} 
    The plastic component of the loss modulus, $G''_p(\gamma_0)$, is plotted as a function of the strain amplitude normalized by peak position in strain amplitude $\gamma_M$ for several values of the plasticity production exponent $\lambda$, as indicated in the legend. The elastoviscoplastic parameters are identical to those utilized in Fig.~\ref{fig:viscoplastic_decomposition}. Solid lines overlaid on the data represent power-law fits of the form $G''_p = G_p^0 \gamma_0^{\xi}$ performed within the pre-yielding linear regime (defined by the condition $G''_p < G''_v$). \textbf{Inset:} Fitted power-law exponents $\nu$ extracted from the numerical data as a function of $\lambda$. The solid line denotes the function $\xi = 2/(2-\lambda)$, demonstrating excellent agreement across the entire investigated parameter range.}
    \label{fig:plastic_scaling_exponents}  
    \label{fig:plastic2}
\end{figure}

A particularly remarkable feature of the proposed BFF model concerns the scaling behavior of the plastic loss component, $G''_p(\gamma_0)$, in the pre-yielding regime. As illustrated in Fig.~\ref{fig:plastic2}, we find a power-law decay of the plastic contribution to the first-harmonic loss modulus for decreasing amplitude. The associated scaling exponent is uniquely dictated by the plasticity production exponent, $\lambda$, systematically obeying the scaling relation:
\begin{equation}
    G''_p(\gamma_0) \sim \gamma_0^{\xi}= \gamma_0^{\frac{2}{2-\lambda}}.
\end{equation}
We obtained the same scaling for any other choice of parameters in the ranges $15$ s$^{-1}$ $\leq r\leq 150000$ s$^{-1}$, $300$ Pa$ \leq G_c \leq 30000$ Pa, $0.067$ s$^{-1}$ $\leq \Gamma_0 \leq 0.00067$ s$^{-1}$, $300$ Pa$ \leq \sigma_c \leq 30000$ Pa.
The exponent obtained for the same model parameters as in figure \ref{fig:viscoplastic_decomposition} is shown in the inset of Fig.~\ref{fig:plastic2}. The pre-yielding non-linear fingerprint is thus an intrinsic topological property governed solely by the plasticity production kinetics.

\newpage

\section{Bifurcation diagrams}
\begin{figure}[htbp]
    \centering
    \includegraphics[width=1.0\linewidth]{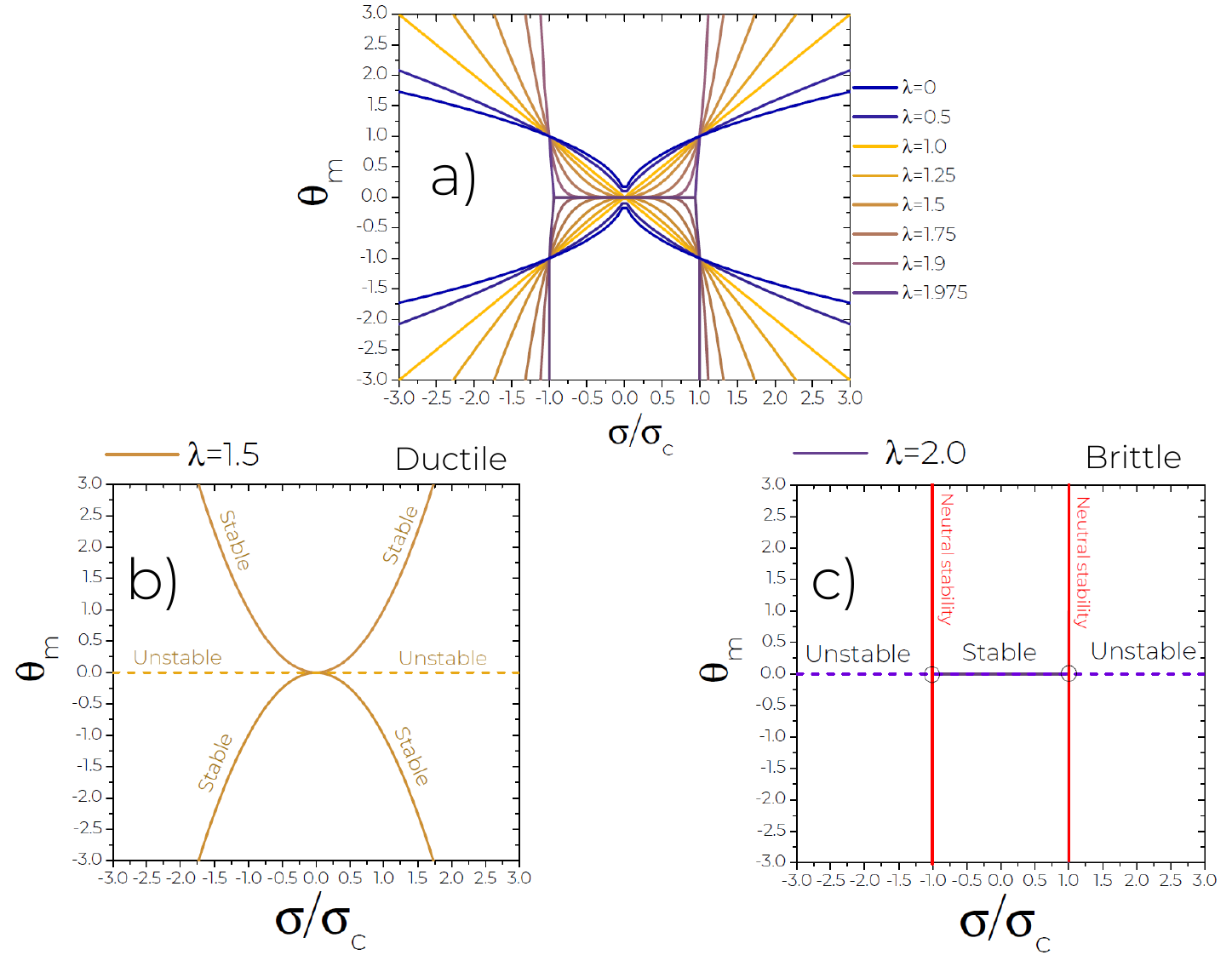}
    \caption{\textbf{Bifurcation diagrams and stability analysis of the effective potential $V(\theta)$.} 
    The curves track the positions of the fixed points $\theta_m$ as a function of the normalized stress $\sigma/\sigma_c$. Solid lines denote stable fixed points (minima of the potential), while dashed lines represent unstable fixed points (maxima of the potential). 
    \textbf{(a)} Evolution of the bifurcation diagram upon varying the plasticity production exponent $\lambda$ from $0$ to $1.975$, showing the progressive opening of the stable branches. 
    \textbf{(b)} Representative ductile response for $\lambda = 1.5$, where the system exhibits a smooth, continuous transition from the unstable state ($\theta_m = 0$) to the stable branches upon increasing the stress. 
    \textbf{(c)} Brittle response at the critical threshold $\lambda = 2.0$. The diagram features a central stable region for $\theta_{m}=0$ flanked by vertical boundaries at $\sigma/\sigma_c = \pm 1$ representing lines of neutral stability (where the second derivative of the potential vanishes, $\partial^2 V / \partial \theta^2 = 0$), separating the stable core from the unstable outer regime.}  
    \label{fig:bifurcation_diagrams}
\end{figure}

To uncover the mathematical origin of the ductile-to-brittle transition in the BFF model, we analyze the fixed points of the dynamical system governed by the effective potential $\mathcal{V}(\theta) = \frac{1}{2}\alpha\theta^2 - |\sigma\theta|^\lambda$, where the stationary points $\theta_m$ are determined by the condition $\partial \mathcal{V} / \partial \theta = 0$. From a dynamical systems perspective, the parameter $\lambda$ acts as a structural control variable that profoundly alters the topology of the phase space, as illustrated in the bifurcation diagrams of Fig.~\ref{fig:bifurcation_diagrams}a).

In the regime $0 < \lambda < 2$, the system undergoes a continuous, supercritical pitchfork-like bifurcation at $\sigma = 0$ (Fig.~\ref{fig:bifurcation_diagrams}a,b). For $\sigma=0$, the origin $\theta_m = 0$ represents the sole stable minimum of the potential. Upon increasing $\sigma$, the central fixed point undergoes a continuous loss of stability, giving rise to two symmetric, stable branches that smoothly emerge from zero. This continuous state-selection describes a gradual plastic yielding, which macroscopically manifests as a ductile rheological response.

Conversely, at the critical threshold $\lambda = 2$, the nature of the transition undergoes a radical mutation towards a discontinuous, brittle-like collapse (Fig.~\ref{fig:bifurcation_diagrams}c). At this specific exponent, the coupling between the stress and the order-like parameter $\theta$ becomes strictly quadratic, and the potential reduces to $\mathcal{V}(\theta) = \frac{1}{2}(\alpha - 2\sigma^2)\theta^2$. Consequently, the system no longer exhibits a standard localized bifurcation. Instead, it undergoes a global stability crossover: for $|\sigma| < \sigma_c = \sqrt{\alpha/2}$, the origin remains strictly stable, whereas for $|\sigma| > \sigma_c$ the potential instantly flips, rendering the origin unstable without generating any neighboring stable fixed points. Exactly at the critical boundaries $\sigma = \pm \sigma_c$, the second derivative of the potential identically vanishes ($\partial^2 \mathcal{V} / \partial \theta^2 = 0$), generating vertical lines of neutral stability where the entire potential landscape becomes flat. The point $\lambda = 2$ thus defines a global bifurcation threshold separating continuous, Landau-like state selections from abrupt, structural destabilizations.


\newpage

\newpage

\section{Entropy production rate $\dot{S}_0/\dot{S}_{max}^{(1)}$ of fig 4-f in linear scale}

\begin{figure}[htbp]
    \centering
    \includegraphics[width=1.0\linewidth]{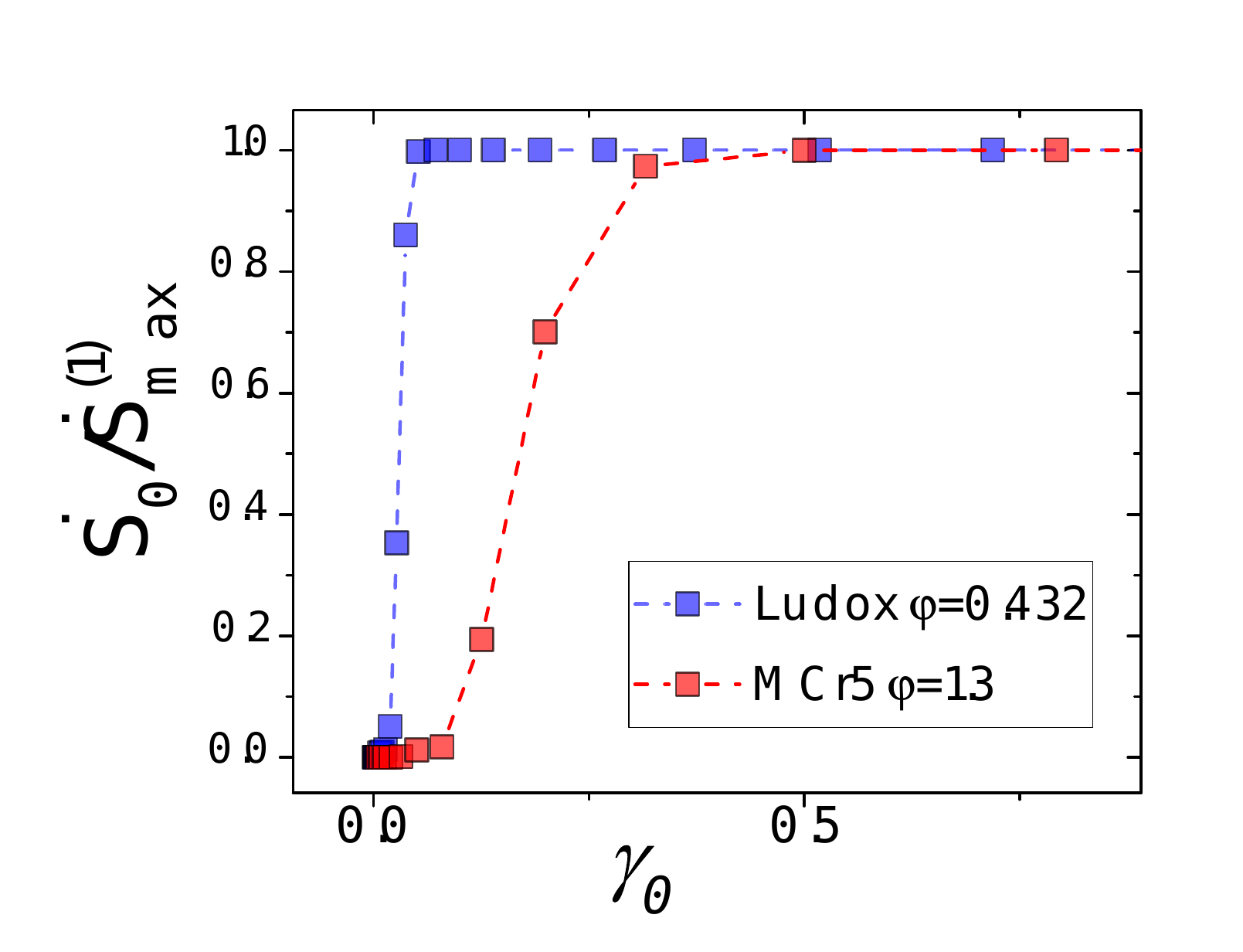}
    \caption{Same entropy production ratios shown in figure 4e-f of the main manuscript in linear scale for both axes.}  
    \label{max_entropy_prod_0}
\end{figure}

\bibliography{Universal_Scaling_LAOS}